\documentclass{aastex62}
\usepackage{bm}
\usepackage{amsmath, empheq, amssymb}
\usepackage{graphicx}

\newcommand{\partialb}{\operatorname{\bm{\nabla}}_\mathrm{\bm b = \bm{0}}}
\newcommand{\intq}{\int d^2 \bm q}
\newcommand{\intr}{\int d^2 \bm r}
\newcommand{\Q}{Q(\bm q)}

\newcommand{\e}{\operatorname{e}}
\newcommand{\fv}{f_V(\bm r; \bm b, \lambda)}

\newcommand{\etarms}{\langle |\bm{\eta}_\mathrm{0, a} (\lambda)| ^2 \rangle}
\newcommand{\fvq}{\tilde{f}_V(\bm q, \lambda)}
\newcommand{\ft}{\operatorname e^{-i \bm q \cdot \bm r}}
\newcommand{\grad}{\bm \nabla} 
\newcommand{\Iea}{I_\mathrm{\mathrm{ea}} (\bm r)}
\newcommand{\qdl}{\frac{\bm q}{D \lambdabar}}
\newcommand{\powerlaw}{\left |\frac{\bm r}{r_\mathrm{\mathrm{diff}}} \right|^\alpha}
\newcommand{\bq}{\bm q}
\newcommand{\bb}{\bm b}

\newcommand{\ftb}{\e^{-i \bm r \cdot \bm b/(D \lambdabar)}}
\newcommand{\etacov}{\langle \bm \eta_\mathrm{0, a}(\bm r, \lambda_1) \cdot \bm \eta_\mathrm{0, a}(\bm r + \Delta \bm r, \lambda_2) \rangle}
\newcommand{\fluct}{\bm \eta_\mathrm{0, a} (\lambda)}
\newcommand{\fiw}{\bm{f}_\mathrm{iw} (\bm r, \lambda)}
\newcommand{\fiwone}{\bm{f}_\mathrm{iw} (\bm r_1, \lambda)}
\newcommand{\fiwtwo}{\bm{f}_\mathrm{iw} (\bm r_2, \lambda)} 
\newcommand{\fiwq}{\tilde{\bm f}_\mathrm{iw} (\bq, \lambda)}
\newcommand{\deltaeta}{ \langle \left[\bm\eta_\mathrm{0, a}(\bm r + \Delta \bm r, \lambda_1) - \bm \eta_\mathrm{0, a}(\bm r, \lambda_2) \right]^2 \rangle}

\newcommand{\Detalambdarms}{\langle \left|\Delta \bm \eta_\mathrm{0, a} (\Delta \bm r; \lambda_1, \lambda_2) \right|^2 \rangle}
\newcommand{\Detarms}{\langle \left|\Delta \bm \eta_\mathrm{0, a} (\Delta \bm r, \lambda) \right|^2 \rangle}
\newcommand{\Deta}{\Delta \bm{\eta}_\mathrm{0, a} (\Delta \bm r, \lambda)}

\def\lsim{\mathrel{\raise.3ex\hbox{$<$\kern-.75em\lower1ex\hbox{$\sim$}}}}
\def\gsim{\mathrel{\raise.3ex\hbox{$>$\kern-.75em\lower1ex\hbox{$\sim$}}}}
\def\gtwid{\mathrel{\raise.3ex\hbox{$>$\kern-.75em\lower1ex\hbox{$\sim$}}}}
\def\proptwid{\mathrel{\raise.3ex\hbox{$\propto$\kern-.75em\lower1ex\hbox{$\sim$}}}}

\graphicspath{{./}{figures_pdf/}}

\shorttitle{Limitations from Scattering for Testing GR}
\shortauthors{Zhu et al.}

\begin{document}

\title{Testing General Relativity with the Black Hole Shadow Size and Asymmetry of Sagittarius A*: Limitations from Interstellar Scattering}

\correspondingauthor{Ziyan Zhu}
\email{zzhu1@g.harvard.edu}

\author{Ziyan Zhu}
\affil{Harvard-Smithsonian Center for Astrophysics, 60 Garden Street, Cambridge, MA 02138, USA}

\author{Michael D. Johnson}
\affiliation{Harvard-Smithsonian Center for Astrophysics, 60 Garden Street, Cambridge, MA 02138, USA}

\author{Ramesh Narayan}
\affiliation{Harvard-Smithsonian Center for Astrophysics, 60 Garden Street, Cambridge, MA 02138, USA}
 
\begin{abstract}

The Event Horizon Telescope (EHT), a global 230\,GHz VLBI array, achieves an angular resolution of ${\approx} 20\,\mu{\rm as}$, sufficient to resolve the supermassive black hole Sagittarius A* (Sgr~A*). This resolution may soon enable measurement of the black hole ``shadow'' size and asymmetry, predicted to be ${\approx}50\,\mu$as and ${\lsim} 3\,\mu$as, respectively. Measurements that depart from these values could indicate a violation of the ``no-hair theorem.'' However, refractive scattering by the turbulent ionized interstellar medium distorts the image of Sgr~A*, affecting its apparent size and asymmetry. In this paper, we present a general analytic approach to quantify the expected image wander, distortion, and asymmetry from refractive scattering. If the turbulence in the scattering material of Sgr~A* is close to Kolmogorov, we estimate the mean refractive image wander, distortion, and asymmetry to be 0.53\,$\mu$as, 0.72\,$\mu$as, and 0.52\,$\mu$as at 230\,GHz. However, alternative scattering models with flatter power spectra can yield larger values, up to 2.1\,$\mu$as, 6.3\,$\mu$as, and 5.0\,$\mu$as, respectively. We demonstrate that these effects can be reduced by averaging images over multiple observations. For a small number of observations, the effects of scattering can be comparable to or greater than those from black hole spin, and they determine a fundamental limit for testing general relativity via images of Sgr~A*. 
\end{abstract}


\section{Introduction} \label{sec:intro}

In general relativity (GR), astrophysical black holes are described by the Kerr metric, parameterized entirely by their mass $M$ and spin $a$ (the so-called ``no-hair theorem'') \citep{israel1967event, israel1968event,carter1971axisymmetric,carter1973black,hawking1972black, robinson1975uniqueness}. The severe light bending near a black hole leads to enhanced emission at the photon ring, which encircles the black hole ``shadow'' \citep{bardeen1973black, luminet1979image,Falcke_2000}.  For a Kerr black hole, the azimuthally-averaged half opening angle of the shadow is predicted to lie in the range $(5 \pm 0.2)\Theta_M$, where $\Theta_M \equiv GM/Dc^2$ and $D$ is its distance from the Earth \citep[e.g.,][]{psaltis2015general}. The uncertainty of the half opening angle arises from the variation of the spin and inclination. In addition, the shadow of a Kerr black hole is nearly circular, with only modest asymmetry even at large spin \citep{johannsen2010testing}. Thus, measuring the size and asymmetry of a black hole shadow would provide a null hypothesis test of GR \citep{psaltis2015general}.

A prime candidate for such a test is the Galactic Center supermassive black hole, Sagittarius A* (Sgr~A*), which has a mass $M \sim 4 \times 10^6 M_\odot$ and lies at a distance $D = 8.1\,$kpc \citep{schodel2003stellar, ghez2008measuring, reid2014trigonometric, boehle2016improved, gillessen2017update, 2018A&A...615L..15G}. Thus, for Sgr~A*, one Schwarzchild radius subtends $2\Theta_M \approx 10\,\mu$as, and GR predicts the asymmetry of Sgr~A* to be less than $0.6\Theta_M$ or $3\,\mu$as \citep[e.g.,][]{johannsen2010testing, chan2013gray}. These scales are comparable to the angular resolution of very-long-baseline interferometry (VLBI) with the Event Horizon Telescope \citep[EHT;][]{doeleman2009imaging}, which may soon generate images of Sgr~A*  \citep[see, e.g.,][]{chael2016high,galaxies4040054,Lu_2016,johnson2016stochastic,Johnson_2017,Bouman_2017,Chael_Closure}. 

However, Sgr~A* images at radio wavelengths are strongly affected by scattering in the ionized interstellar medium (ISM). Scattering introduces image blurring, substructure, wander, and distortion \citep{blandford1985low,narayan1989shape,goodman1989shape,johnson2015theory}. Although scattering mitigation techniques have been proposed \citep{fish2014imaging, johnson2016stochastic}, no analytic framework has been developed to quantify how scattering will affect tests of GR or alternate theories of gravity using images of Sgr~A*.  

In this paper, we quantify the image wander and distortion due to interstellar scattering by extending a framework originally developed by \citet{blandford1985low}. We begin in Section \ref{sec:background} by reviewing some of the basic principles of scattering and introduce important mathematical tools. Next, in Section \ref{sec:theory}, we present our framework to calculate image wander and distortion from scattering and compare our results with numerical simulations of scattering. In Section \ref{sec:results}, we apply our analytic framework to estimate the mean image wander, distortion, and asymmetry from scattering at 230 and 345\,GHz for two proposed scattering models of Sgr~A*, and we discuss the implications for testing GR with the EHT. We summarize our results in Section \ref{sec:conclusion}. Additional technical details are given in the Appendix.


\section{Background} \label{sec:background}
In this section, we summarize the basic framework for interstellar scattering of radio waves and introduce the necessary mathematical tools to compute refractive image wander and distortion. More detailed discussion of the background theory can be found in \citet{blandford1985low}, \citet{rickett1990radio}, \citet{narayan1992physics}, and \citet{thompson2001interferometry}. 

\begin{figure*}[ht!]
\centering
\includegraphics[width = 0.55\linewidth]{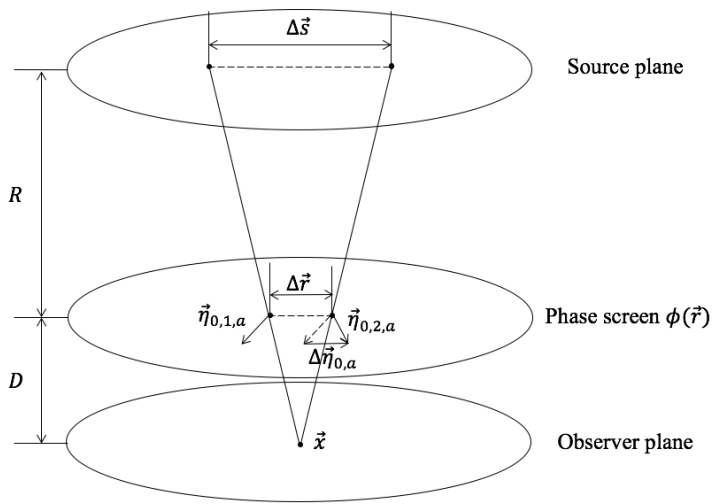}
\caption{The setup of the thin screen model for quantifying image distortion. A phase screen $\phi(\bm r)$ is located at a distance $D$ from the observer and $R$ from the source. Two sources separated by a distance $\Delta \bm s$ on the source plane can be projected onto a separation $\Delta \bm r$ on the phase screen by multiplying $\Delta \bm s$ by a geometric factor $D/(D+R)$; two observers separated by $\Delta \bm x$ on the observer plane can also be projected onto the phase screen by multiplying $\Delta \bm x$ by $R/(D+R)$. The two point sources separated by $\Delta \bm r$ on the phase screen each have a different image centroid shift vector $\bm \eta_\mathrm{0, i, a}, \ i = 1, 2$. $\Delta \bm \eta_\mathrm{0, a} = \bm \eta_\mathrm{0, 1, a} - \bm \eta_\mathrm{0, 2, a}$. Assuming point 2 is located at the origin, $\Delta \bm \eta_\mathrm{0, a}$ is the image wander of point 1 relative to point 2, which we define to be the distortion of point 1 (both magnitude and direction).}\label{fig:geometry}
\end{figure*}

\subsection{Refraction in the Turbulent Interstellar Medium}

Refractive scattering by the turbulent ISM at radio wavelengths is caused by density inhomogeneities in the ionized ISM \citep{scheuer1968amplitude}. The local index of refraction of the ISM is given by $n \approx 1-\frac{1}{2}\left(\nu_p/\nu\right)^2$, where $\nu$ is the wave frequency, $\nu_\mathrm{p}=\sqrt{4 \pi n_\mathrm{e}^2/m_\mathrm{e}}$ is the plasma frequency, and $n_e$ is the local electron density. A density fluctuation $\delta n_e$ over a path length $dz$ introduces a phase fluctuation $\delta \phi = - r_e \lambda \times dz \times \delta n_e$, where $r_e = \e^2/(m_e c^2) \approx 2.8 \times 10^{-13}$ cm is the classical electron radius. In many cases, the density fluctuations can be well-described as a turbulent cascade with a power law power spectrum \citep{rickett1977interstellar}. For theoretical convenience, the turbulence may be approximated to be confined in a thin screen $\phi (\bm r)$, located a distance $D$ from the observer and $R$ from the source. The corresponding magnification is defined to be $M = D/R$. Figure \ref{fig:geometry} shows the basic setup of the thin screen model. In this figure and throughout the paper, $\bm s$, $\bm r$, and $\bm x$ denote transverse vectors on the source plane, phase screen, and the observer plane respectively. As radio waves propagate through the turbulent ISM, transverse gradients in the screen phase cause the incident waves to change propagation direction. In the remainder of the paper, the screen is assumed to be ``frozen,'' meaning that the scattering evolution is deterministic and only depends on the relative transverse motions of the observer, screen, and the source with a characteristic velocity $ \mathbf{V} _\mathrm{\perp}$. 

Three important length scales describe the scattering by a thin phase screen: the diffractive scale $r_\mathrm{\mathrm{diff}}$, the refractive scale $r_\mathrm{ref}$, and the Fresnel scale $r_\mathrm{F}$. The screen phase statistics are described via $r_\mathrm{\mathrm{diff}}$. It corresponds to the transverse length on the phase screen over which the RMS phase difference is 1 rad. The diffractive scale is related to the ensemble-average scatter-broadening angle of a point source $\theta_\mathrm{scatt}$ by $r_\mathrm{\mathrm{diff}} =  2 \sqrt{2 \ln 2} \lambdabar/[\theta_\mathrm{scatt} (D/R+1)]$, where $\lambdabar \equiv \lambda/(2 \pi)$. Diffractive effects are those that arise from modes on scales comparable to $r_\mathrm{diff}$. The refractive scale $r_\mathrm{ref}$ is the projected size of angular broadening on the scattering screen, $r_\mathrm{ref} = \theta_\mathrm{scatt} D$. Refractive effects are those that are dominated by modes on scales comparable to $r_\mathrm{ref}$. The Fresnel scale $r_\mathrm{F}$ corresponds to the geometric mean of the diffractive and refractive scales, $r_\mathrm{F} = \sqrt{r_\mathrm{ref}r_\mathrm{diff}}$. It is completely defined in terms of the geometrical parameters of the scattering by $r_\mathrm{F} \equiv \sqrt{DR\lambdabar/(D+R)}$. Physically, $r_\mathrm{F}$ is the transverse distance on the phase screen over which the geometrical path difference between two rays is roughly $\lambda$. This paper focuses on the strong scattering limit, defined by the regime where $r_\mathrm{\mathrm{F}} \gg r_\mathrm{\mathrm{diff}}$ \citep[e.g.,][]{blandford1985low}. 

\cite{narayan1989shape} and \cite{goodman1989shape} showed that there are three distinct image averaging regimes in the strong scattering limit: ``snapshot'' image, ``average'' image, and ``ensemble average'' image. The snapshot image corresponds to averaging over timescales less than $r_\mathrm{\mathrm{diff}}/\mathbf{V}_\mathrm{\perp}$. A snapshot image exhibits variability due to both refractive and diffractive effects and can be interpreted as an instantaneous snapshot. The average image averages over timescale less than $r_\mathrm{ref}/\mathbf{V}_\mathrm{\perp}$ and only exhibits variability from refractive scattering, while diffractive effects cause a time-independent blurring. The ensemble-average image corresponds to averaging over infinite time. It is equivalent to convolving the original image by a blurring kernel $G(\bm r)$. Throughout the paper, subscripts ``ss'', ``a'', and ``ea'' denote snapshot, average, and ensemble-average images respectively.  For a source larger than $r_\mathrm{diff}/D$, diffractive effects are quenched. This condition is met at all radio wavelengths for Sgr~A* (e.g., $r_\mathrm{diff}/D \approx 0.008\, \mathrm{\mu as}$ for Sgr~A* at $1.3 \, \mathrm{mm}$). Therefore, we focus on average and ensemble average regimes in this paper. 

\subsection{Statistical Properties of the Phase Screen}\label{sec:stats}

The statistical properties of the phase screen $\phi (\bm r)$ can be described in two complimentary ways. The first approach is the structure function, which is defined to be the following, 
\begin{equation}
D_\mathrm{\phi} (\bm r) \equiv \left \langle [ \phi(\bm r' + \bm r) - \phi(\bm r') ]^2 \right \rangle \propto \lambda^2.
\label{eqn:dphi}
\end{equation}
Here and throughout the paper, $\langle ... \rangle$ denotes an ensemble average of realizations of the screen phase, which can be approximated in practice by a time average. The second approach to describe the statistical character of the phase screen is the power spectrum of the phase fluctuations, $Q(\bq)$. Typically, the power spectrum can be characterized by a single power law between some inner scale $r_\mathrm{in}$ and outer scale $r_\mathrm{out}$: $Q(\bq) \propto |\bq|^{-\beta}$, where $\bq$ is a two-dimensional wave vector and $\beta$ is the power law index. For example, for Kolmogorov turbulence, $\beta = 11/3$ \citep{goldreich1995toward}. Defining the two-point correlation, 
\begin{equation}
C(\bm r) \equiv \langle \phi (\bm r' + \bm r) \phi (\bm r)\rangle,
\end{equation}
the power spectrum is the Fourier transform of $C(\bm r)$, 
\begin{equation}
Q(\bq) = \frac{1}{\lambdabar^2}\intr\ C(\bm r) \e^{-i \bq \cdot \bm r}.
\label{eqn:Qq_structure}\end{equation}
Since $\phi (\bm r) \propto \lambda$ (Equation (\ref{eqn:dphi})), $Q(\bq)$ is a dimensionless quantity that is independent of wavelength. Also noting that $D_\phi = 2[C(0) - C(\bm r)]$, Equation (\ref{eqn:Qq_structure}) can be expressed as the following, 
\begin{equation}
Q(\bq) \equiv -\frac{1}{2 \lambdabar^2}\tilde{D}_\phi(\bq),
\label{eqn:Qq_of_Dphi}
\end{equation}
where the tilde denotes a two-dimensional Fourier transform, adopting the following convention throughout the paper, 
\begin{equation}
\tilde{f} (\bq) = \intr f(\bm r) \e^{-i \bq \cdot \bm r}.
\end{equation}

In the ensemble-average scattering limit, the scattering simply acts to convolve the unscattered image with a blurring kernel $G(\bm r)$: 
\begin{equation}
I_\mathrm{ea} (\bm r) = I_\mathrm{src} (\bm r) \star G(\bm r),
\label{eqn:kernel}
\end{equation}
where $\star$ denotes a spatial convolution. The blurring kernel is most naturally represented in the Fourier domain, in which the visibility at an interferometric baseline $\bb$ is defined to be the Fourier transform of the intensity. In terms of the structure function of the phase screen $D_\phi$, the blurring kernel at a baseline $\bb$ is \citep[see][]{coles1987refractive}
\begin{equation}
\tilde{G} (\bm b) = \e^{-\frac{1}{2}D_\phi (\bb/(1+M)) }.
\label{eqn:kernel_ft}
\end{equation}
For spatial displacements shorter than $r_\mathrm{in}$, the scale on which turbulence is dissipated, the phase fluctuations $\phi (\bm r)$ vary smoothly. Therefore, in this limit, $\phi(\bm r' + \bm r) \approx \phi (\bm r') + \bm r \cdot \grad \phi (\bm r')$, and $D_\mathrm{\phi} \propto \lambda^2|\bm r|^2$ \citep{tatarskii1971effects}. This shows that the ensemble average broadening acts like a Gaussian blurring kernel with FWHM $\theta_\mathrm{scatt} \propto \lambda^2$ for baselines $|\bm b| \lesssim (1+M) r_\mathrm{in}.$ For displacements greater than $r_\mathrm{in}$, the phase structure function varies as a power law in $\bm r$ over the inertial range of the turbulence, until saturating at $r_\mathrm{out}$. For $|\bm r| \in [r_\mathrm{in}, r_\mathrm{out}]$, or equivalently, $|\bb| \in \left[(1+M)r_\mathrm{in}, (1+M)r_\mathrm{out}\right]$, the structure function scales with $\bm r$ as $D_\phi (\bm r) \propto \lambda^2 |\bm r|^\alpha$, where $\alpha = \beta - 2$. In this range, the scattering kernel becomes non-Gaussian and the angular broadening angle scales as $\theta_\mathrm{scatt}\propto \lambda^{1+2/\alpha}$. The outer scale $r_\mathrm{out}$ is physically associated with the scale on which the turbulence is injected and is usually much larger than any of the other scales. Interstellar scattering is often observed to be anisotropic, and we denote the major and minor axes sizes of the scattering broadening kernel as $\theta_\mathrm{maj}$ and $\theta_\mathrm{min}$, respectively.

Refractive effects on images can be approximated through the gradient of the phase screen $\grad \phi (\bm r)$. For a given realization of the phase screen $\phi (\bm r)$, the average image is given as follows \citep[see Equations (9) and (10) in][]{johnson2016optics}, 
\begin{align}
I_\mathrm{a} (\bm r) &\approx I_\mathrm{\mathrm{ea}} (\bm r + r_\mathrm{\mathrm{F}}^2 \grad \phi (\bm r)) \nonumber \\
& = I_\mathrm{src} \star G(\bm r) +r_\mathrm{F}^2[\grad \phi (\bm r ) \cdot \grad(I_\mathrm{src} (\bm r) \star G(\bm r))].
\label{eqn:i_approx}
\end{align}
Here we have used Equation (\ref{eqn:kernel}) to write the ensemble-average image as the spatial convolution of the intrinsic source $I_\mathrm{src} (\bm r)$ with the diffractive blurring kernel $G (\bm r)$. The second term describes the image distortion due to refraction. The relations above allow us to derive a mathematical tool to express the mean and correlation of normalized fluctuations, which we will employ in the remainder of the paper. 

\subsection{Covariance between Two Scintillating Quantities}

Consider the fluctuation caused from scattering $\delta A (\lambda)$ of an observable quantity, such as total flux density or image centroid. Suppose that $\delta A (\lambda)$ can be written in the following general form, 
\begin{equation}
\delta A(\lambda) = \intr \ \phi(\bm r, \lambda) f(\bm r, \lambda),
\label{eqn:fluct_general}
\end{equation}
where $\phi (\bm r, \lambda)$ is the screen phase at position $\bm r$, and $f (\bm r, \lambda)$ is a complex function associated with the fluctuation $\delta A (\lambda)$. The covariance between two fluctuating quantities $\delta A_1 (\lambda_1)$ and $\delta A_2 (\lambda_2)$ can then be obtained as follows, 
\begin{align}
\langle \delta A_1(\lambda_1) \delta A_2 (\lambda_2) \rangle = &\left\langle \intr_1 \intr_2\ \phi(\bm r_1, \lambda_1) \phi(\bm r_2, \lambda_2) f_1 (\bm r_1, \lambda_1) f_2 (\bm r_2, \lambda_2) \right\rangle \nonumber \\
= & \frac{\lambdabar_1 \lambdabar_2}{(2\pi)^2} \intq\ Q(\bq) \tilde{f_1}(-\bq, \lambda_1) \tilde{f_2} (\bq, \lambda_1),
\label{eqn:corr_general}
\end{align}
where the second equality follows from Equation (\ref{eqn:Qq_structure}), which is equivalent to
\begin{equation}
\langle \phi(\bm r_1) \phi(\bm r_2) \rangle = \frac{\lambdabar^2}{(2 \pi)^2} \int d^2 \bm q\ Q(\bm q) \e^{i \bm q \cdot (\bm r_1 - \bm r_2)}.
\end{equation}

More generally, if the phase screen is moving at some velocity, and $\delta A_1$ and $\delta A_2$ are observed at different times, or if the sources or the observers of $\delta A_1$ and $\delta A_2$ are separated by some distance, the covariance between them is
\begin{align}
\langle \delta A_1( \bm r, \lambda_1) \delta A_2 (\bm r + \Delta \bm r, \lambda_2) \rangle=& \left \langle \intr_1 \intr_2\ \phi(\bm r_1, \lambda_1) \phi(\bm r_2 + \Delta \bm r, \lambda_2) f_1 (\bm r_1, \lambda_1) f_2 (\bm r_2, \lambda_2) \right \rangle \nonumber \\
=& \frac{\lambdabar_1\lambdabar_2}{(2\pi)^2} \intq \ Q(\bq) \tilde{f_1}(-\bq, \lambda_1) \tilde{f_2} (\bq, \lambda_1) \e^{-i \bq \cdot \Delta \bm r},
\label{eqn:corr_dr}
\end{align}
where $\Delta \bm r$ is the projected separation between $\delta A_1$ and $\delta A_2$ on the phase screen. If the phase screen moves at a velocity $\mathbf{V}_\perp$ and $\delta A_2$ is observed at a time $t$ after $\delta A_1$, $\Delta \bm r = \mathbf{V}_\perp t$. If the sources are separated by $\Delta \bm s$ on the source plane, a separation $\Delta \bm s$ on the source plane can be projected onto the phase screen by $\Delta \bm r = D\bm \Delta s/(D+R) $. Similarly, a separation $\Delta \bm x$ on the observer's plane is projected onto the phase screen by $\Delta \bm x = R \bm \Delta r/(D+R)$ (see Figure \ref{fig:geometry}). Note that $\Delta \bm s$, $\Delta \bm r$, and $\Delta \bm x$ have the same values in angular units. In this paper, we mainly consider the case in which the two sources are located at different positions. 

\begin{figure}[ht!]
\centering
	\includegraphics[width = \linewidth]{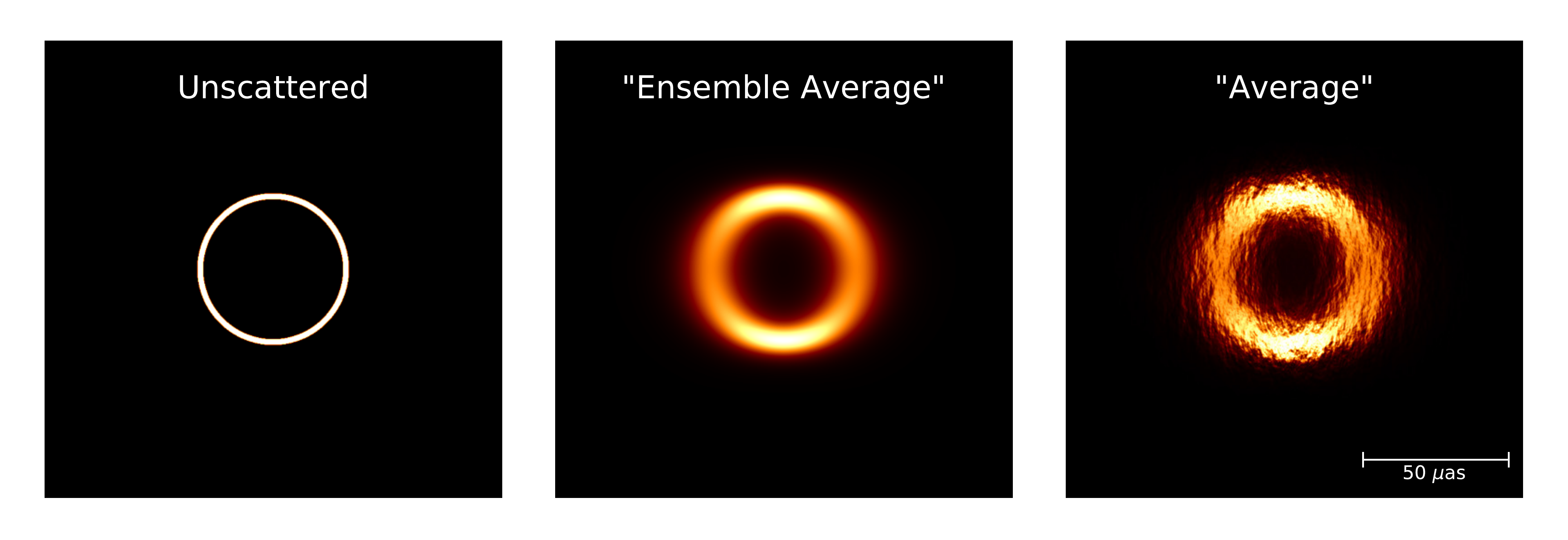}
	\caption{An example showing a uniform ring, its ensemble average image, and a realization of a scattered image. The unscattered model (left) is a ring with source size $\theta_\mathrm{src} = 52\,\mu$as and thickness of 5.2$\,\mu$as. Ensemble average scattering causes blurring of the original image (middle), while the average image (right) exhibits both blurring and substructure due to refractive scattering. The scattering parameters correspond to those of Sgr~A* at $1.3\, \mathrm{mm}$ using the J18 model. }
\label{fig:photon_ring}
\end{figure}

\subsection{Scattering Parameters of Sgr~A*}{\label{sec:scattering_params}}

In this section, we discuss the scattering properties of Sgr~A* as determined by \citet{johnson2018scattering}. The location of the phase screen has been measured through the temporal broadening of the Galactic Center magnetar $2.4''$ from Sgr~A* \citep{bower2015proper}: $D = 2.7\,$kpc and $R = 5.4\,$kpc. The blurring kernel $G(\bm r)$ is anisotropic. To compute its FWHM major and minor axes, we identify the dimensionless baseline length such that $\exp \left[-\frac{1}{2} D_\phi (\bm u_{1/2} (1 + M)) \right] \equiv 1/2$, and then use the relation $\theta_\mathrm{FWHM} = \frac{2 \ln 2}{\pi u_\mathrm{1/2}}$ to find FWHM major and minor axes. At a position angle 90$^\circ$, the FWHM major and minor axes of the scattering kernel are $\theta_\mathrm{maj} \approx 23\, \mathrm{\mu as}$ and $\theta_\mathrm{min} \approx 12\, \mathrm{\mu as}$ at $230 \, \mathrm{GHz}$, and they are $\theta_\mathrm{maj} \approx 10\, \mathrm{\mu as}$ and $\theta_\mathrm{min} \approx 5\, \mathrm{\mu as}$ at $345 \, \mathrm{GHz}$.

We consider two different scattering models in our calculations. The primary model \citep{johnson2018scattering} has $r_\mathrm{in} = 800\,\mathrm{km}$ and $\alpha = 1.38$, the latter being slightly shallower than a Kolmogorov spectrum ($\alpha = 5/3$). We refer to this as the J18 model. A second scattering model is inspired by \citet{goldreich2006folded}, who model the ISM scattering as a collection of folded current sheets \citep{schekochihin2004simulations}. This model reproduces the $\lambda^2$ scaling of angular size at centimeter wavelengths as well as Gaussian scatter-broadening for Sgr~A*, and corresponds to $\alpha = 0$. To be consistent with refractive visibility measurements in Sgr~A* at $1.3\, \mathrm{cm}$ and $3.5 \, \mathrm{cm}$, this model requires the inner scale to be $r_\mathrm{in} \approx 2\times10^6\, \mathrm{km}$ \citep[see Figure 14 in][]{johnson2018scattering}. We call this the GS06 model. In both models, we assume $r_\mathrm{out}$ is sufficiently large that it is irrelevant for our calculations. We take the intrinsic source size of Sgr~A* to be $\theta_\mathrm{src} \approx 400\lambda_\mathrm{cm} \,\mathrm{\mu as} $.

\subsection{Numerical Simulations: Stochastic Optics}\label{sec:stochastic}

The simulations in this work are performed with the \texttt{stochastic-optics} module from the \texttt{eht-imaging} Python package \citep{chael2016high, johnson2016stochastic}. Here, we briefly summarize some general considerations of the stochastic optics model and describe how it is implemented. A more detailed description can be found in \citet{johnson2016stochastic}.

In this model, the unscattered image $I_\mathrm{src} (\bm r)$, the blurring kernel $G(\bm r)$, and the time-averaged power spectrum $Q (\bq)$ are given, while the refractive phase screen is treated as stochastic. The code generates random realizations of $\phi (\bm r)$ and their corresponding scattered average image $I_\mathrm{a} (\bm r)$. The phase screen $\phi (\bm r)$ is represented in the frequency domain, $\tilde{\phi} (\bq) = \intr \phi(\bm r) \e^{-i \bq \cdot \bm r}$, and the Fourier components $\tilde{\phi} (\bq)$ are uncorrelated, complex Gaussian random variables with the mean square amplitude $Q(\bq)$ and a random phase. The scattering phase screen $\tilde{\phi}$ is discretized using an $N \times N$ grid of Fourier coefficients $\tilde{\phi}_{s, t}$:
\begin{align}
\phi_{l,m} &= \frac{1}{F^2} \sum^{N-1}_{s, t = 0} \tilde{\phi}_{s, t} \e^{2\pi i (ls + mt)/N} \nonumber \\
&=\frac{\lambdabar}{F} \sum^{N-1}_\mathrm{s, t = 0} \sqrt{Q(s,t)} \times \epsilon_{s, t} \e^{2\pi i (ls+mt)/N},
\label{eqn:philm}
\end{align}
where $F$ is the field of view expressed as a transverse length on the phase screen, and $F/N$ is the image resolution. $\epsilon_{s, t}$ is a set of Gaussian complex variables defined to be $\epsilon_{s,t} \equiv \tilde{\phi}_{s, t}/\sqrt{Q(s,t)}$. We set the mean of $\epsilon_{s, t}$ to be zero ($\epsilon_{0, 0} = 0$), which is an arbitrary choice because the average phase does not affect the image. Note that $\epsilon_{s, t}$ is wavelength and model independent. To ensure $\phi_{l,m}\in \mathbb{R}$, we require that $\epsilon_{s, t} = \epsilon^{*}_{-s, -t}$. Therefore, the phase screen is represented by $N_\mathrm{\phi} \equiv (N^2 - 1)/2$ independent parameters. 

\begin{figure}[ht!]
\centering
	\includegraphics[width = \linewidth]{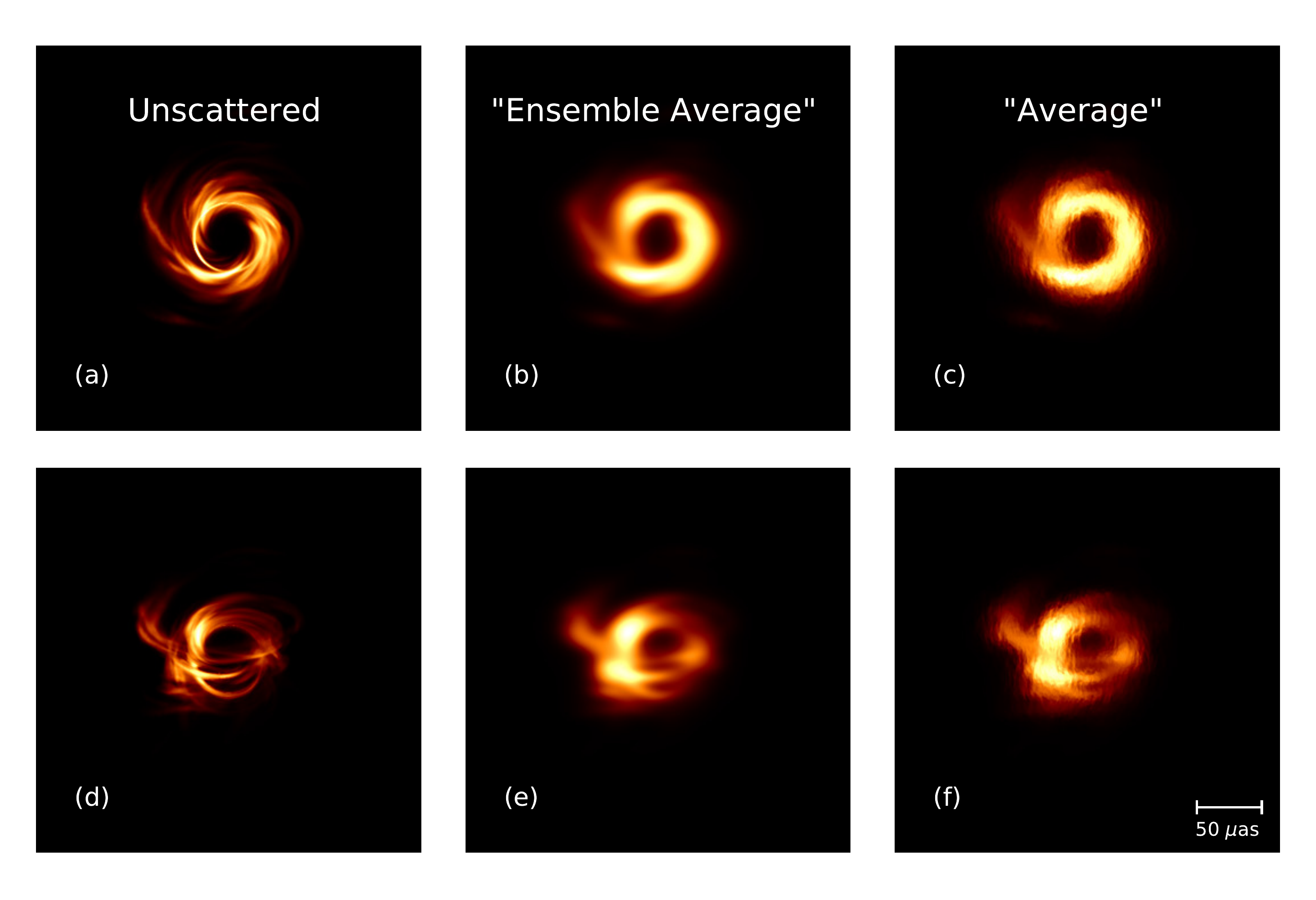}
    \caption{Left: Two 230 GHz images from GRRMHD simulations of Sgr~A* at $10^{\circ}$ (top) and $60^{\circ}$ (bottom) inclination angles \citep[see Figures 11 and 12 in][]{chael2018role}. Middle: Ensemble average images showing image blurring. Right: Average images with scattering model J18 showing refractive substructure.}
    \label{fig:scattered_sgrA}
\end{figure}

Figures \ref{fig:photon_ring} and \ref{fig:scattered_sgrA} show two examples of scattered images with the J18 model. Figure \ref{fig:photon_ring} shows unscattered and scattered images of a ring of size 52$\,\mu$as, which is the expected ring diameter of Sgr~A*. Figure \ref{fig:scattered_sgrA} shows two frames from two-temperature
general relativistic radiative magnetohydrodynamic (GRRMHD) simulations of Sgr~A* \citep{chael2018role} performed with the code \texttt{KORAL} \citep{Sadowski_2013,Sadowski_2014,Sadowski_2017} as well as the corresponding scattered images. Ensemble average images exhibit blurring effects, and refractive substructure is visible in all presented scattered images.

\section{Image Wander and Distortion} \label{sec:theory}
\subsection{Refractive Fluctuations of the Image Centroid}
In this section, we quantify the image centroid shift due to refractive noise (image wander) using the mathematical tools derived in Section \ref{sec:stats}. The image centroid $\bm{r}_\mathrm{0, a}$ of an average image is defined as
\begin{equation} 
\bm{r}_\mathrm{0, a} \equiv D \bm \eta_\mathrm{0, a} \equiv \frac{\intr \, \bm{r} I_\mathrm{a} (\bm{r})\ }{\int d^2 \bm{r}\, I_\mathrm{a}(\bm{r})},
\end{equation}
where $I_\mathrm{a} (\bm r)$ is the intensity as a function of projected location $\bm r$ on the scattering screen, and $\bm \eta_\mathrm{0, a}$ is its centroid in angular units. With the assumption that the ensemble-average image is centered at the origin, $\bm r_\mathrm{0, ea} = \bm 0$, $\bm r_\mathrm{0, a}$ is the spatial image wander and $\bm \eta_\mathrm{0, a}$ is the angular image wander. Image wander can also be expressed in terms of visibility $V_\mathrm{a} (\bb)$, the Fourier transform of $I_\mathrm{a}(\bm{r})$ at a baseline $\bm b$ \citep{thompson2001interferometry}, 
\begin{equation} 
V_\mathrm{a} (\bm{b}) = \intr \, I_\mathrm{a}(\bm r) \operatorname{e}^{- i \bm{b} \cdot \bm{r}/(\lambdabar D)}.
\end{equation} 
The total flux density is $\int d^2 \bm{r} \, I_\mathrm{a}(\bm r) = V_\mathrm{a} (\bm{0})$. Taking the gradient of the visibility at zero baseline, $\grad V_\mathrm{a} (\bm{b}) |_\mathrm{\bm{b} = \bm{0}} = - \frac{i}{\lambdabar D} \int d^2 \bm{r} \, \bm{r} I(\bm{r})$ and denoting $ \operatorname{\bm{\nabla}} V_\mathrm{a} (\bm{b}) |_\mathrm{\bm{b} = \bm{0}} \equiv \operatorname{\bm{\nabla}}_\mathrm{\bm{b} = \bm{0}} V_\mathrm{a} (\bm{b})$, the angular image centroid $\bm \eta_\mathrm{0, a}$ can be rewritten in terms of the gradient of the visibility at zero baseline,
\begin{equation}
\fluct = i \lambdabar \frac{\partialb V_\mathrm{a} (\bm{b})}{V_\mathrm{a} (\bm{0})}.
\label{eqn:centroid}
\end{equation}

To calculate refractive fluctuations in the image centroid, we need to separate refractive fluctuations from the ensemble-average image. The visibility of the average image $V_\mathrm{a} (\bm{b})$ is the sum of the ensemble average visibility $V_\mathrm{ea} (\bm{b})$ and the refractive fluctuation $\Delta V_\mathrm{a} (\bm{b})$, i.e., 
\begin{equation}
V_\mathrm{a} (\bm b) = V_\mathrm{ea} (\bm b) + \Delta V_\mathrm{a} (\bm b).
\end{equation}
Assuming the refractive fluctuation term $\Delta V_\mathrm{a} (\bm{b})$ is  much smaller than $V_\mathrm{a} (\bm{b})$ in the vicinity of $\bm b = \bm 0$, we can Taylor expand $\fluct$ to linear order in $\Delta V_\mathrm{a} (\bm{0})/V_\mathrm{ea}(\bm 0)$, 
\begin{align}
\fluct &\approx \frac{i \lambdabar}{V_\mathrm{ea} (\bm{0})} \left [1 - \frac{\Delta V_\mathrm{a}(\bm 0)}{V_\mathrm{ea}(\bm 0)} \right] \operatorname{\bm{\nabla}}_\mathrm{\bm{b} = \bm{0}} [V_\mathrm{ea} (\bm b) + \Delta V_\mathrm{a} (\bm b)] \nonumber \\
& = \frac{i \lambdabar}{V_\mathrm{ea} (\bm 0)} \operatorname{\bm{\nabla}}_\mathrm{\bm{b} = \bm{0}} \Delta V_\mathrm{a} (\bm b),
\label{eqn:image_wander_general}
\end{align}
where we have taken the ensemble-average image to be centered on the origin, i.e., $\operatorname{\bm{\nabla}}_\mathrm{\bm{b} = \bm{0}} V_\mathrm{ea} (\bm b) = 0$.

Equation (\ref{eqn:image_wander_general}) is a general expression for the refractive fluctuation of the image centroid in terms of the refractive fluctuation of the visibility of the average image, $\Delta V_a (\bb)$. We can proceed by writing down an explicit form for $\Delta V_a (\bb)$ in terms of $\phi (\bm r)$. Using Equation (\ref{eqn:i_approx}), the visibility of the average image on a baseline $\bb$ can be rewritten as follows \citep[see Equation (11) from][]{johnson2016optics},
\begin{align}
V_\mathrm{a}(\bb) & \approx V_\mathrm{ea} (\bb) + \intr \,  r_\mathrm{\mathrm{F}}^2 \grad \phi(\bm r) \cdot \grad I_\mathrm{\mathrm{ea}} (\bm r) \ftb \nonumber \\
& = V_\mathrm{ea} (\bb) + r_\mathrm{\mathrm{F}}^2 \intr \, \e^{-i \bm b \cdot \bm r/(D \lambdabar)} \left[ \frac{i}{D \lambdabar} \bm b \cdot \grad \Iea - \grad^2 \Iea \right] \phi(\bm r),
\end{align} 
where the last equality is obtained by integration by parts. The refractive fluctuation $\Delta V_a (\bb)$ can be extracted from the equation above \citep[see Equation (12) from][]{johnson2016optics}:
\begin{equation} 
\Delta V_\mathrm{a} (\bm b) = \int d^2 \bm r \, f_V (\bm r; \bm b, \lambda) \phi(\bm r, \lambda)
\label{eqn:fV}
,\end{equation}
where 
\begin{equation}
\fv = r_\mathrm{\mathrm{F}}^2 \e^{-i \bm b \cdot \bm r/(D \lambdabar)} \left[ \frac{i}{D \lambdabar} \bm b \cdot \grad \Iea - \grad^2 \Iea \right]
\label{eqn:fv}
\end{equation}
is the function that maps to the visibility fluctuation. It is an example of $f (\bm r, \lambda)$ in Equation (\ref{eqn:fluct_general}) that characterizes the fluctuation. Plugging Equation (\ref{eqn:fV}) into Equation (\ref{eqn:image_wander_general}), we obtain the refractive image wander in angular units 
\begin{align}
\bm\eta_\mathrm{0, a} (\lambda) &=  \frac{i \lambdabar}{V_\mathrm{ea} (\bm 0)}  \int d^2 \bm r \, [\partialb f_V (\bm r; \bm b, \lambda)] \phi(\bm r, \lambda)\nonumber \\
& \equiv \intr\, \fiw \phi(\bm r),
\label{eqn:x0a}
\end{align}
where 
\begin{align}
\fiw& \equiv \frac{i\lambdabar}{V_\mathrm{ea} (\bm 0)} \grad_{\bm b = \bm 0} \fv \nonumber \\
&=\frac{i \lambdabar r_\mathrm{\mathrm F}^2}{V_\mathrm{ea}(\bm 0)}\partialb \left\{ \e^{-i \bm b \cdot \bm r/(D \lambdabar)} \left[ \frac{i}{D \lambdabar} \bm b \cdot \grad \Iea - \grad^2 \Iea \right] \right\}
\label{eqn:fiw}
\end{align}
is the function that maps to refractive fluctuations in the image centroid. It can be determined given the source ensemble-average image $I_\mathrm{ea} (\bm r)$, which is obtained by convolving the source image $I_\mathrm{src} (\bm r)$ with $G (\bm r)$ (Equation (\ref{eqn:kernel})). Note that $\fiw$ is a vector quantity; because both $\fluct$ and $\phi (\bm r) \in \mathbb{R}$, $\fiw \in \mathbb{R}^2$. 

\subsection{RMS Fluctuations and Covariance of Image Wander}\label{sec:rms}
The RMS fluctuation of refractive image wander can be obtained from Equations (\ref{eqn:corr_general}) and (\ref{eqn:x0a}):
\begin{align}
\sqrt{\etarms} &= \sqrt{\langle \bm{\eta}_\mathrm{0, a}(\lambda) \cdot \bm{\eta}_\mathrm{0, a}(\lambda)} \rangle \nonumber \\
& = \frac{\lambdabar}{2\pi} \sqrt{ \intq \,\Q |\fiwq|^2 },
\label{eqn:xrms}
\end{align} 
where 
\begin{align}
\fiwq &= \intr \, \fiw \ft 
\label{eqn:fiwq}
\end{align}
is the Fourier transform of $\fiw$ given in Equation (\ref{eqn:fiw}). Because $\fiw \in \mathbb{R}^2$, $ \fiwq = \tilde{\bm f}_\mathrm{iw}^* (-\bm q, \lambda)$.

Equation (\ref{eqn:xrms}) provides a closed-form expression to calculate the root-mean-squared refractive image wander given the power spectrum $Q(\bq)$, the scattering kernel $G(\bm r)$, and the unscattered image $I_\mathrm{src} (\bm r)$. As a demonstration, we calculate analytically the image wander for an isotropic Gaussian source and an isotropic Kolmogorov spectrum ($\alpha = 5/3$) in the Appendix. In this case, the image wander scales as $\left(\frac{r_\mathrm{\mathrm{diff}}}{r_\mathrm{F}}\right)^{2-\alpha} \left(\frac{\theta_\mathrm{scatt}}{\theta_\mathrm{ea}}\right)^{1-\alpha/2} \left(\frac{\lambda}{r_\mathrm{\mathrm{diff}}} \right)$, where $\theta_\mathrm{ea}$ is the FWHM of the ensemble-average image, also defined in the Appendix. The magnitude of image wander increases as a function of $\lambda$, whereas most refractive effects decrease as a function of $\lambda$. In reality, the scattering of Sgr~A* is anisotropic and $Q(\bq)$ is more complicated than a single unbroken power law. Therefore, we perform numerical integration to calculate image wander in the rest of the paper, which we will refer to as the semi-analytic model. 

More generally, using Equation (\ref{eqn:corr_dr}), we can calculate the covariance between the image wander of two points separated by $\Delta \bm r$ on the phase screen with observing wavelengths $\lambda_1$ and $\lambda_2$, respectively. We will denote angular image wander for observations at a wavelength $\lambda$ and a position $\bm r$ on the phase screen as $\bm \eta_\mathrm{0, a} (\bm r, \lambda)$. Then, by Equation (\ref{eqn:corr_dr}),
\begin{align} 
\etacov & = \left \langle \intr_1 \intr_2 \, \phi (\bm r_1, \lambda_1) \phi(\bm r_2 + \Delta \bm r, \lambda_2) \fiwone \cdot \fiwtwo \right \rangle \nonumber \\
& = \frac{\lambdabar_1\lambdabar_2}{(2 \pi)^2} \intq\ Q(\bm q) \tilde{\bm f}_\mathrm{iw}^* (\bq, \lambda_1) \cdot \tilde{\bm f}_\mathrm{iw} (\bq, \lambda_2) \e^{-i \bm q \cdot \Delta \bm r}.
\label{eqn:cov}
\end{align} 

\subsection{Image Distortion}\label{sec:distortion}
We define relative image wander between two points separated by $\Delta \bm r$ as follows (also see the definition of $\Delta \bm \eta_\mathrm{0, a}$ in Figure \ref{fig:geometry}),
\begin{equation} 
\Delta \bm \eta_\mathrm{0, a} (\Delta \bm r; \lambda_1, \lambda_2) \equiv \bm \eta_\mathrm{0, a}(\bm r + \Delta \bm r, \lambda_1) - \bm \eta_\mathrm{0, a}(\bm r, \lambda_2).
\end{equation} 
Using this definition, the structure function of the relative wander is, 
\begin{align}
\Detalambdarms \equiv& \deltaeta  \nonumber \\
=& \left(\frac{\lambdabar_1}{2\pi}\right)^2 \intq \, \Q |{\tilde{\bm f}_\mathrm{iw} (\bq, \lambda_1)}|^2+\left(\frac{\lambdabar_2}{2\pi}\right)^2 \intq \, \Q |{\tilde{\bm f}_\mathrm{iw} (\bq, \lambda_2)}|^2 \nonumber \\
&-\frac{\lambdabar_1\lambdabar_2}{(2 \pi)^2} \intq\, Q(\bm q) \left[\tilde{\bm f}_\mathrm{iw}^* (\bq, \lambda_1) \cdot \tilde{\bm f}_\mathrm{iw} (\bq, \lambda_2) +\tilde{\bm f}_\mathrm{iw}^* (\bq, \lambda_2) \cdot \tilde{\bm f}_\mathrm{iw} (\bq, \lambda_1)\right]\e^{-i \bm q \cdot \Delta \bm r}.
\end{align} 

For the purpose of quantifying image distortion, we consider the case where $\lambda_1 = \lambda_2 \equiv \lambda$. The RMS relative position wander from refractive scattering is
\begin{align} 
\sqrt{\Detarms} & = \frac{\lambdabar}{\sqrt{2} \pi} \sqrt{ \intq \, Q(\bq) |\fiwq|^2 (1-\e^{-i \bq \cdot \Delta \bm r})}.
\label{eqn:detarms}
 \end{align}
This expression quantifies the relative image wander between two points separated by $\Delta \bm r$ on the phase screen. Since we can project $\Delta \bm r$ onto the source plane (Figure \ref{fig:geometry}), this is equivalent to calculating the relative image wander between two separated sources. We can then quantify image distortion from scattering by computing wander of any point on the image relative to the wander of the image centroid. Take a uniform ring as an example. Calculating the relative image wander $\Delta \bm \eta_\mathrm{0, a} (\Delta \bm r, \lambda)$ between a position on the ring and the origin (i.e., the image centroid of the unscattered ring) gives the distortion of this position due to refractive effects. Tracing the relative image wander at all positions gives the shape of the distorted ring for one realization, and taking the RMS of relative image wander over realizations gives the expected distortion. Large scattering modes (with wavelengths larger than the transverse image size) can cause large image wander without introducing image distortion. Likewise, relative image wander and image distortion are independent of the assumed image centroid, which is not measured with standard VLBI.

We can also calculate the magnitude of the distortion along an arbitrary direction $\bm{\hat{e}}_i$ by making a projection:
\begin{align}
\sqrt{\langle |\Deta \cdot \bm{\hat{e}}_i|^2 \rangle} &=\frac{ \lambdabar}{\sqrt{2} \pi} \sqrt{\intq \, Q(\bq) |\fiwq \cdot \bm{\hat{e}}_i|^2 (1-\e^{-i \bq \cdot \Delta \bm r})}.
\label{eqn:distortion}
\end{align}
Note that $\sqrt{\langle |\Deta \cdot \bm{\hat{e}}_i|^2 \rangle}$ is dependent on $\bm{\hat{e}}_i$ and is generally anisotropic even for an isotropic power spectrum. 

\section{Image Distortion and Asymmetry of Sgr~A*} \label{sec:results}

In this section, we present results from applying the technique in Section \ref{sec:theory} to quantify the image distortion and asymmetry of Sgr~A*. We first present a calculation of the relative image wander between two Gaussian sources and show that the results agree well with numerical simulations. We then demonstrate how to use the same framework to compute the image distortion and the degree of asymmetry of Sgr~A*, approximated as a uniform ring, for EHT observations. Unless otherwise specified, all the scattering parameters are from the J18 model at $1.3\, \mathrm{mm}$. The power spectrum of phase fluctuations in the scattering screen is the ``dipole model'' defined using the framework of \citet{psaltis2018model}; the exact form of $Q(\bq)$ can be found in Appendix B of \citet{johnson2018scattering}.

\subsection{Relative Image Wander between Gaussian Sources}\label{sec:relative}

\begin{figure*}[ht!]
\gridline{\fig{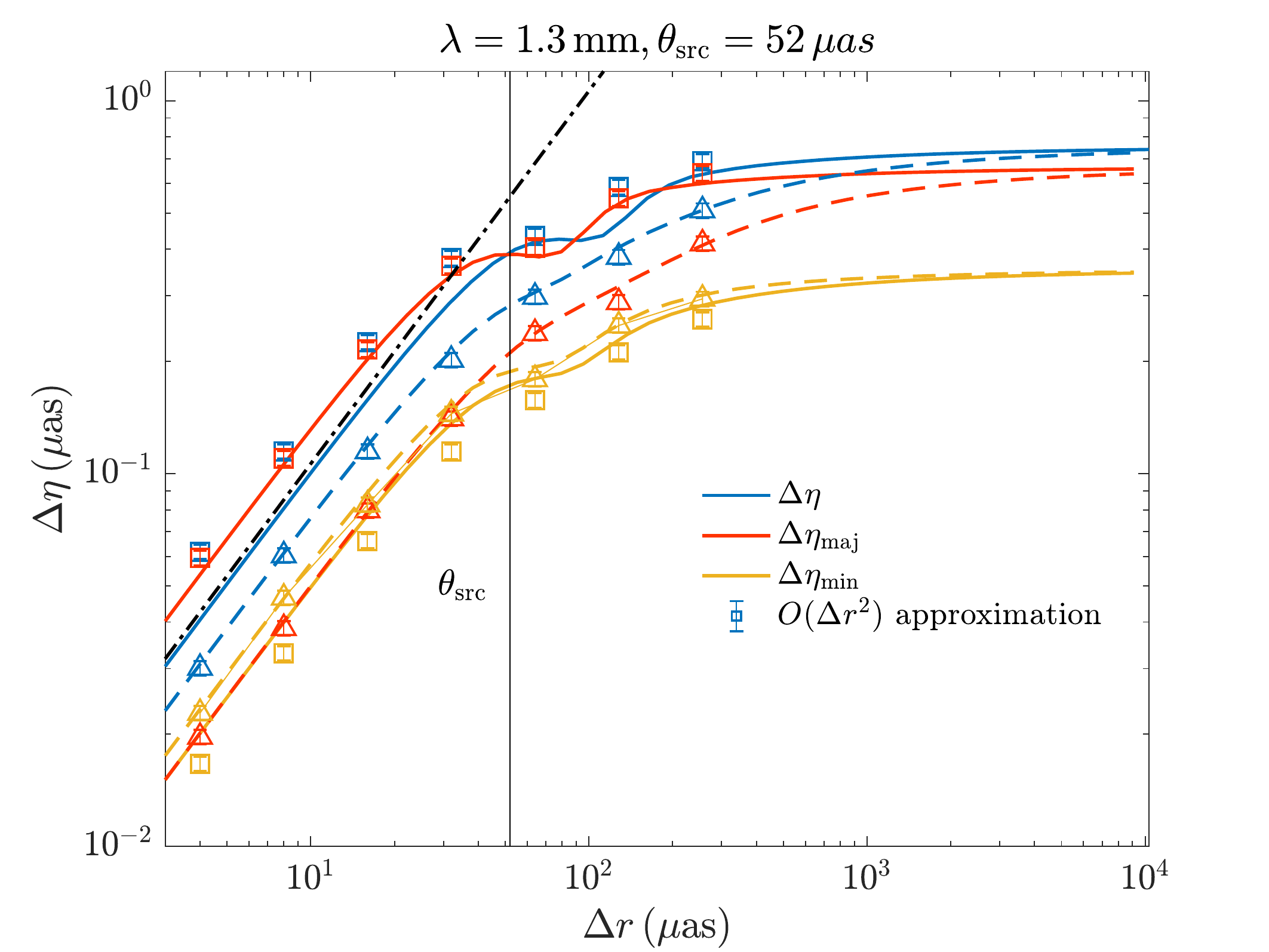}{0.48\textwidth}{(a)}
          \fig{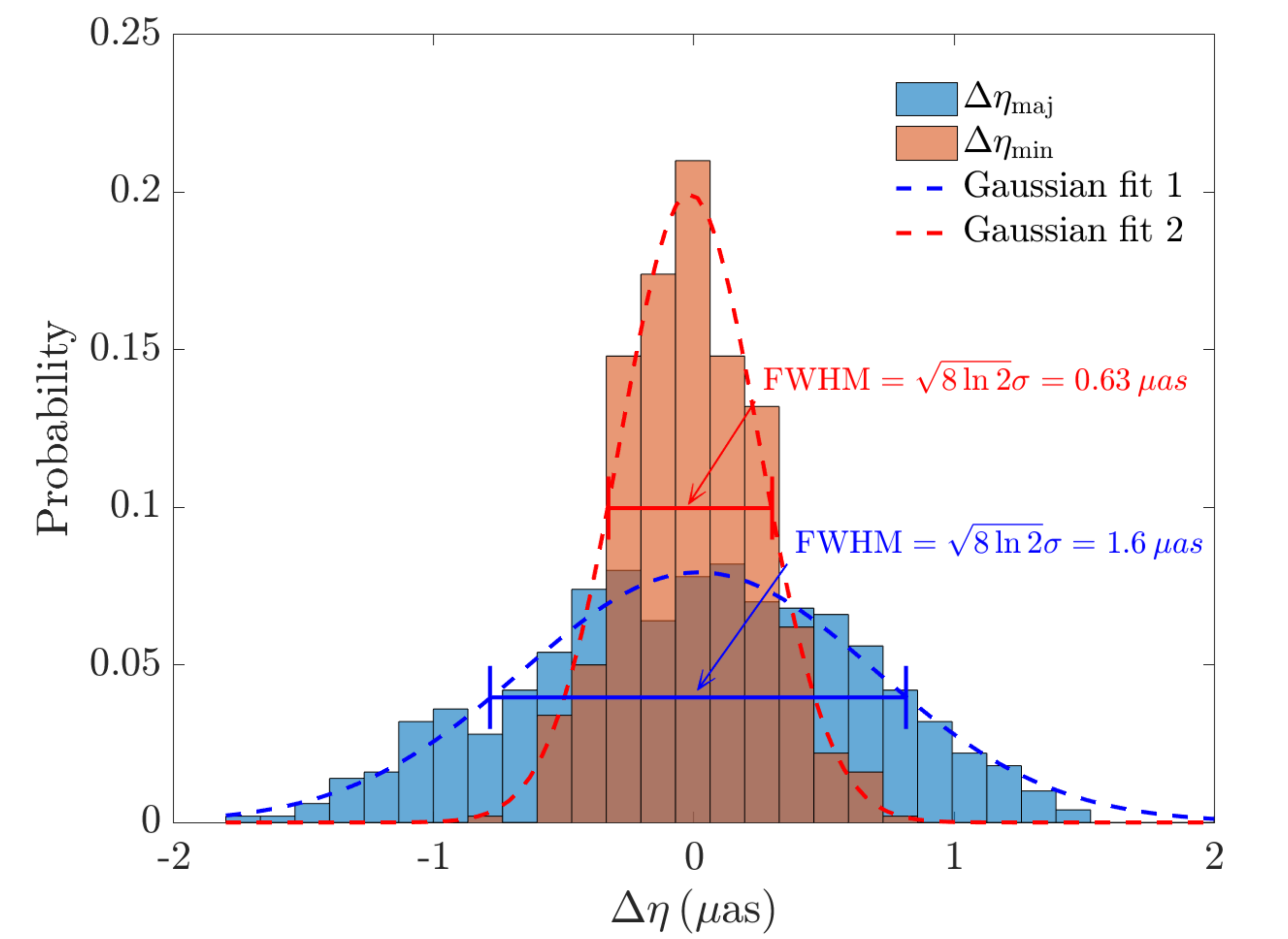}{0.48\textwidth}{(b)}}
    \caption{(a) RMS fluctuations of relative image wander $\Delta \eta$ at 1.3\,mm for the J18 scattering model between two $52\,\mu{\rm as}$ Gaussian sources, as well as its projections along the major ($\Delta \eta_\mathrm{maj}$) and minor ($\Delta \eta_\mathrm{min}$) axes of the scattering kernel. The different line styles represent different displacement directions of the second source. Solid lines: source displacement along $\theta_\mathrm{maj}$; dashed lines: source displacement along $\theta_\mathrm{min}$. The different colors represent different projection directions. Blue represents the magnitude of $\Delta \bm \eta$; red represents the projection along $\theta_\mathrm{maj}$; yellow represents the projection along $\theta_\mathrm{min}$. The black dotted line shows the analytical result for $\sqrt{\langle |\Delta \bm \eta_\mathrm{0, a}|^2 \rangle}$ in the limit of small $|\Delta \bm r|$. Points with errorbars are from numerical simulations. (b) Histogram of values for $\Delta \eta_\mathrm{maj}$ and $\Delta \eta_\mathrm{min}$ from numerical simulations with $|\Delta \bm r| = 256\,\mu$as and $\Delta \bm r$ being aligned with $\theta_\mathrm{maj}$. They are both zero-mean Gaussian functions, and their standard deviations $\sigma$ are the RMS image wander that we calculate in (a), which is related to the full-width-half-maximum of the distribution by $\sigma = \mathrm{FWHM}/2\sqrt{2 \ln 2}$. }
    \label{fig:deta}
\end{figure*}

We first calculate the magnitude of relative image wander projected along different directions between two Gaussian sources with angular size $\theta_\mathrm{src} = 400 \lambda_\mathrm{cm} \, \mathrm{\mu as}$ using Equation (\ref{eqn:distortion}). One source is fixed at the origin, and the second source is located at some displacement vector $\Delta \bm r$. Here, we take $\bm{\hat{e}}_i$ in $\sqrt{\langle |\Deta \cdot \bm{\hat{e}}_i|^2 \rangle}$ to be along the major or the minor axis of the scattering kernel. Figure \ref{fig:deta}(a) shows the magnitude and projections of the relative image wander as a function of displacement $\Delta \bm r$ between two isotropic Gaussian sources with $\theta_\mathrm{src} = 52\, \mu$as at $1.3\,\mathrm{mm}$, the observing wavelength of the EHT. For clarity, we denote the magnitude of relative image wander, its projection along the major and minor axis of the scattering kernel as $\Delta \eta$, $\Delta \eta_\mathrm{maj}$, and $\Delta \eta_\mathrm{min}$ respectively, and they are color-coded to be blue, red, and yellow in the figure. Note that $\Delta \eta = \sqrt{\Delta \eta_\mathrm{maj}^2 + \Delta \eta_\mathrm{min}^2}$. The different line styles (solid and dashed) represent displacements along $\theta_\mathrm{maj}$ and $\theta_\mathrm{min}$ respectively. Points with errorbars are obtained with numerical simulations by averaging over 500 realizations. The simulation shows excellent agreement with the analytic calculation. 

\begin{figure*}[ht!]
\centering
\includegraphics[width = 0.6\linewidth]{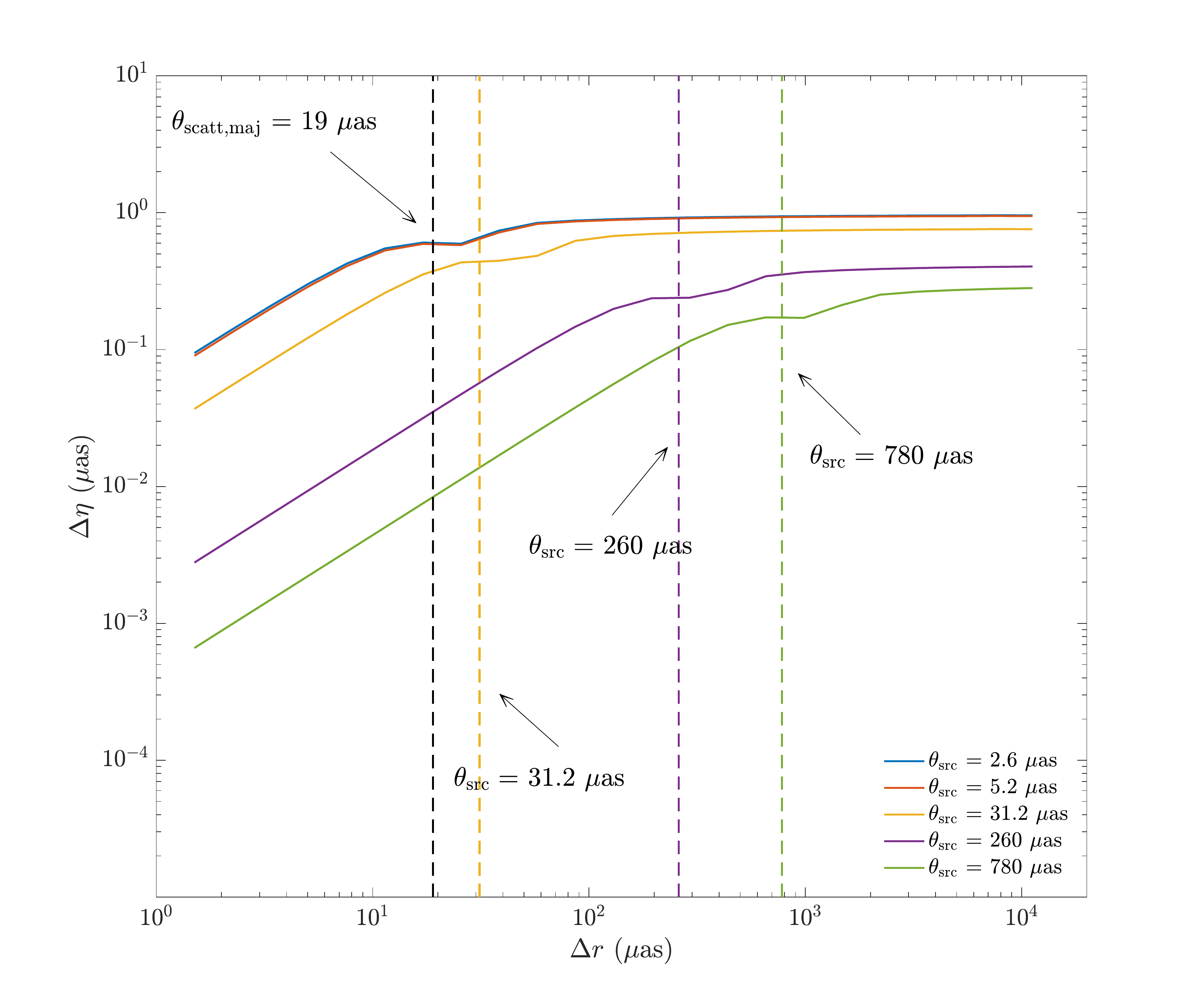}
\caption{Relative image wander at 1.3\,mm for the J18 scattering model as a function of angular separation for different intrinsic source size. Each color corresponds to a different source size, and the magnitude of the source size is marked by the vertical dashed lines. At small displacement, relative wander $\Delta \eta$ is linear in $\Delta r$; at large displacement, it saturates and reaches an asymptotic limit. The threshold displacement that separates these two asymptotic behaviors is $\max[\theta_\mathrm{maj}, \theta_\mathrm{src}]$. The curves from the top to the bottom in the figure correspond to increasing $\theta_\mathrm{src}$. }
\label{fig:deta_source_size}
\end{figure*}

All curves in Figure \ref{fig:deta}(a) exhibit a threshold displacement $\Delta r_\mathrm{thres}$. Below the threshold, $\Delta \eta$ is linear in $|\Delta \bm r|$; above the threshold, $\Delta \bm \eta_\mathrm{0, a}$ saturates and reaches an asymptotic limit. These two distinct behaviors can be understood by considering two limits of $\Delta \bm r$. In the small $\Delta \bm r$ limit, we can expand the exponential in Equation (\ref{eqn:distortion}) to second order in $\Delta \bm r$:
\begin{equation}
\sqrt{\langle |\Deta \cdot \bm{\hat{e}}_i|^2 \rangle} =\frac{ \lambdabar}{\sqrt{2} \pi} \sqrt{\intq \ Q(\bq) |\fiwq \cdot \bm{\hat{e}}_i|^2 \left[i \bq \cdot \Delta \bm r + \frac{1}{2} (\bq \cdot \Delta \bm r)^2 \right]}.
\label{eqn:smallr}
\end{equation}
Since $\sqrt{\langle |\Deta \cdot \bm{\hat{e}}_i|^2 \rangle}$ is purely real, the first order expansion in $\Delta \bm r$ must vanish, and the relative shift becomes
\begin{equation}
\sqrt{\langle |\Deta \cdot \bm{\hat{e}}_i|^2 \rangle} =\frac{ \lambdabar}{2 \pi} \sqrt{\intq \ Q(\bq) |\fiwq \cdot \bm{\hat{e}}_i|^2 (\bq \cdot \Delta \bm r)^2} \propto |\Delta \bm r|, 
\label{eqn:smallr}
\end{equation}
which is linear in $|\Delta \bm r|$. The black dotted line in Figure \ref{fig:deta} represents the small $\Delta \bm r$ expansion of $\Delta \eta$ to second order, which agrees with the exact result up to the threshold displacement $\Delta r_\mathrm{thres}$. In the opposite limit, the two sources are widely separated, and the image wander of each is independent. Therefore, the relative image wander approaches an asymptotic limit that is independent of the displacement direction. 

Figure \ref{fig:deta}(b) shows the distribution of $\Delta \eta_\mathrm{maj}$ and $\Delta \eta_\mathrm{min}$ at $|\Delta \bm r| = 256\,\mu$as and when $\Delta \bm r$ is along the major axis of the scattering kernel $\theta_\mathrm{maj}$. Both distributions are Gaussian as expected because the refractive position wander is a weighted sum of correlated Gaussian random variables $\phi (\bm r)$ (see Equation (\ref{eqn:x0a})) \citep{johnson2016optics}.

Figure \ref{fig:deta_source_size} compares $\Delta \eta$ for various source sizes $\theta_\mathrm{src}$. At small source sizes, the $\Delta \eta$'s almost overlap with each other, and the $\Delta \bm r_\mathrm{thres}$ is the major axis of the scattering broadening angle $\theta_\mathrm{maj}$. Once the source size is larger than $\theta_\mathrm{maj}$, the $\Delta \bm r_\mathrm{thres}$ is given by $\theta_\mathrm{src}$. In a more compact notation, $\Delta \bm r_\mathrm{thres} = \max[\theta_\mathrm{maj}, \theta_\mathrm{src}]$. 

\begin{figure}[ht!]
\centering
	\includegraphics[width = 0.5\linewidth]{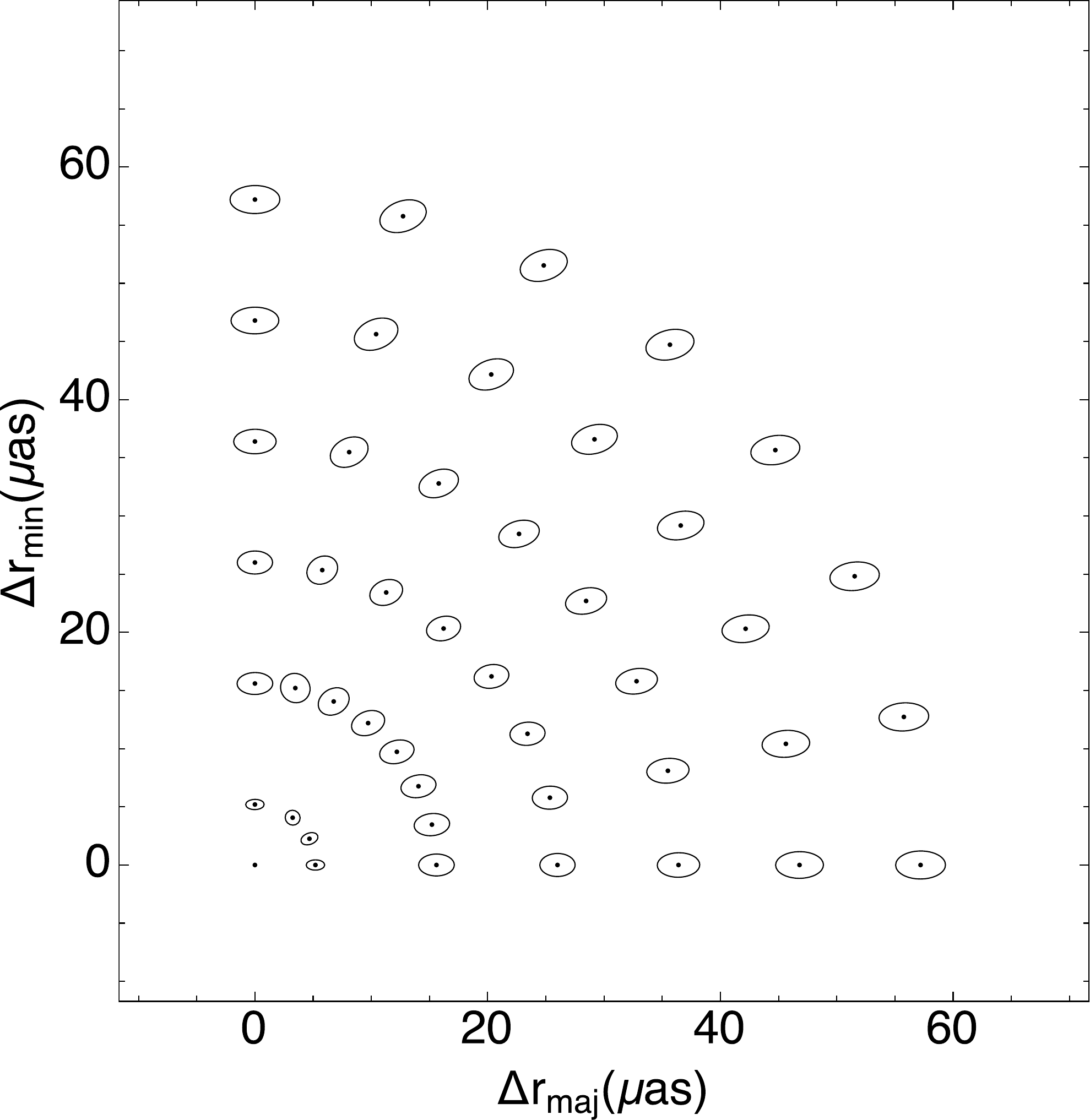}
    \caption{Relative image centroid shift between two Gaussian sources with $\theta_\mathrm{src} = 2 \,\mu$as for various displacements. The ellipses represent the magnitude of the relative centroid shift at a given $\Delta \bm r$, indicated by the black dot at which the ellipses are centered. The ellipse size at each angle shows the RMS position wander projected along that direction. Each ellipse has a magnification of 2.5 for clarity and is aligned along the direction in which the distortion has the largest magnitude. The observing wavelength is $1.3\, \mathrm{mm}$, and the scattering parameters are from the J18 model.}
    \label{fig:graphics}
\end{figure}

We also examine the projection of relative image wander along arbitrary directions. Figure \ref{fig:graphics} shows a visual representation of the relative image wander at various displacement angles between two Gaussian sources, with $\theta_\mathrm{src} = 2\, \mu \mathrm{as}$. Each ellipse represents the relative centroid shift at that position. The orientations of the ellipses are aligned with the direction of the maximum shift. Since the scattering is the strongest along the major axis, the relative wander $\Delta \bm \eta_\mathrm{0, a}$ is expected to be stronger when the second source is displaced along $\theta_\mathrm{maj}$ than along $\theta_\mathrm{min}$. The relative wander is strong also when its projection is along the displacement direction. The combination of these two factors explains why the major axes of the ellipses are aligned between $\theta_\mathrm{maj}$ and the displacement direction for displacement angle between 0 and $\pi/2$ in Figure \ref{fig:graphics}.

\subsection{Refractive Distortion of the Black Hole Shadow of Sgr~A*}

Our formalism to calculate relative image wander from scattering can be used to quantify image distortions from scattering. Such distortions contribute systematic uncertainty to measurements of the image shape, and this uncertainty is independent of practical limitations in image reconstructions (e.g., from finite observing sensitivity or limited baseline coverage). Because we are primarily concerned with estimating refractive distortion of the black hole shadow, we proceed by representing the image of Sgr~A* as a uniform ring with radius $26\,\mathrm{\mu as}$ and a thickness $2\, \mathrm{\mu as}$. We can then evaluate the image distortion at each position on the ring by calculating the relative centroid shift between two Gaussian sources with size equal to the ring thickness, one located at the center of the ring and the other located at the given position on the ring. For any particular scattering realization, the distortion for nearby points on the ring will then be highly correlated. Note that the choice of the ring thickness (i.e., the size of the Gaussian sources) does not significantly modify the shape of the distorted ring unless it exceeds $\theta_\mathrm{maj}$. This is because $\Delta \eta$ is almost independent of $\theta_\mathrm{src}$ when $\theta_\mathrm{src} < \theta_\mathrm{scatt}$ (see Figure~\ref{fig:deta_source_size}, the two curves corresponding to $\theta_\mathrm{src} = 2.6\, \mu$as and $\theta_\mathrm{src} = 5.2 \, \mu$as overlap). 

\begin{figure*}[ht!]
\centering
\gridline{\fig{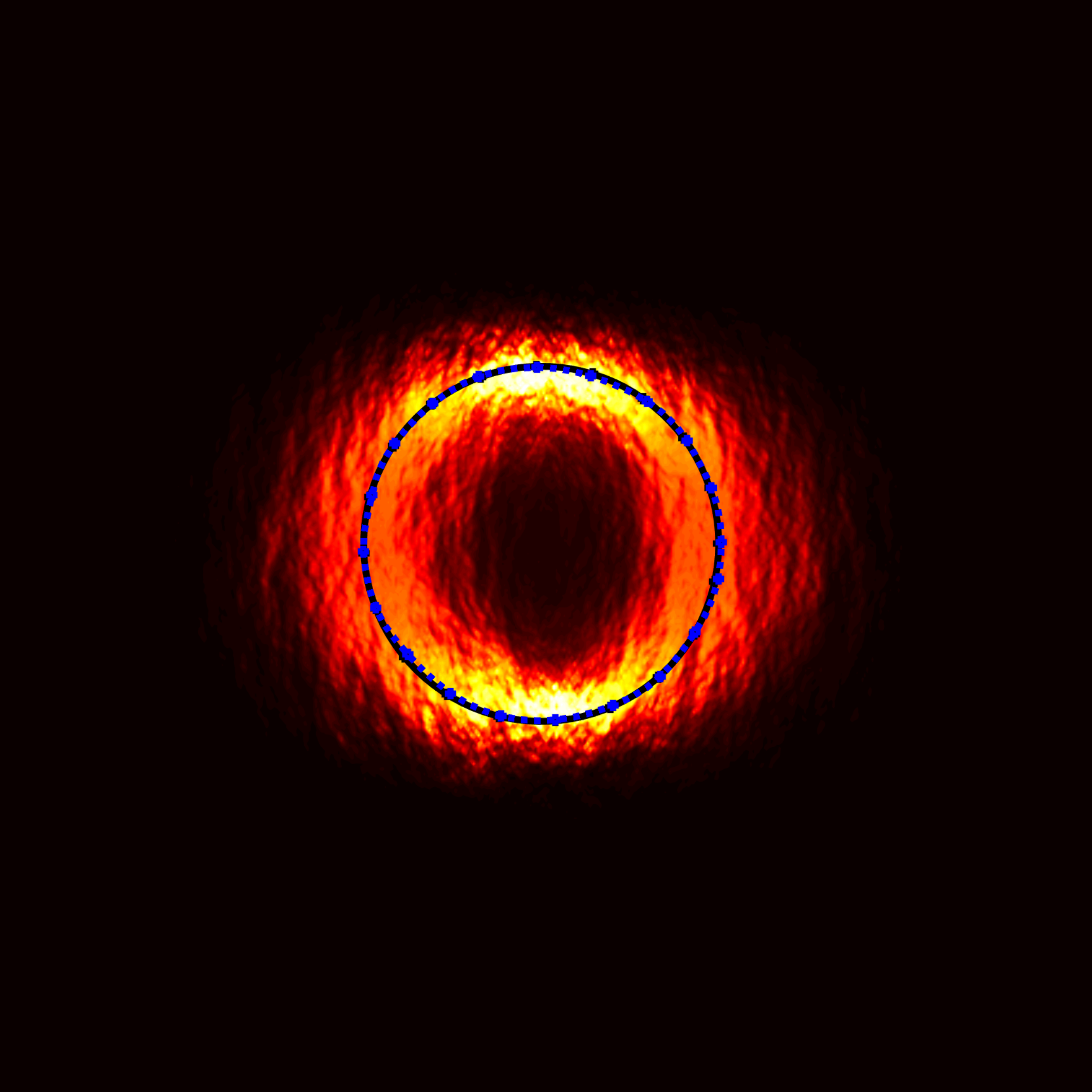}{0.48\textwidth}{(a)}
         \fig{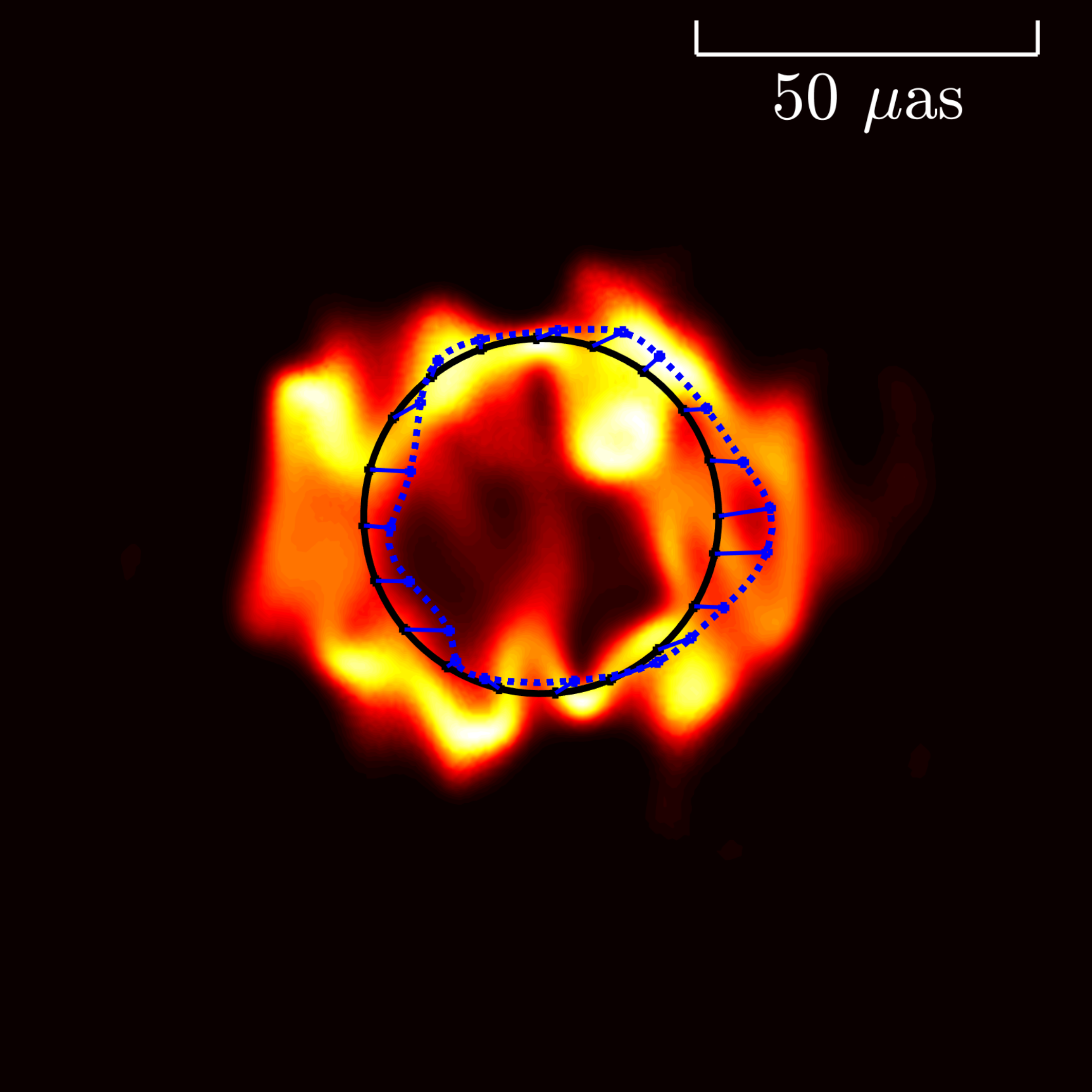}{0.48\textwidth}{(b)}}
    \caption{Examples showing scattered images of a ring with two sets of scattering parameters. Panel (a) adopts the J18 model with $\alpha = 1.38$ and $r_\mathrm{\mathrm{in}} = 800\, \mathrm{km}$; panel (b) generalizes the scattering parameters from the GS06 model with $r_\mathrm{\mathrm{in}} = 2\times 10^6 \, \mathrm{km}$, $\alpha = 0$. We take $\alpha = 0.001$ in the simulation to avoid numerical errors. The black ring has the same diameter as that of the unscattered image but is centered at scattered image centroid (which is not the center of the image). Each blue dot is the shifted image centroid of a scattered Gaussian point source with $\theta_\mathrm{src} = 2\, \mathrm{\mu as}$ located at the corresponding black dot connected by a solid blue line. Tracing a total of 60 blue dots gives the distorted shape (only 20 points are shown). In panel (a), the blue and the black curves almost overlap, whereas in panel (b), there is a significant shift between the two curves.}
\label{fig:ring_distortion}
\end{figure*}

\begin{figure*}[ht!]
\centering
\gridline{\fig{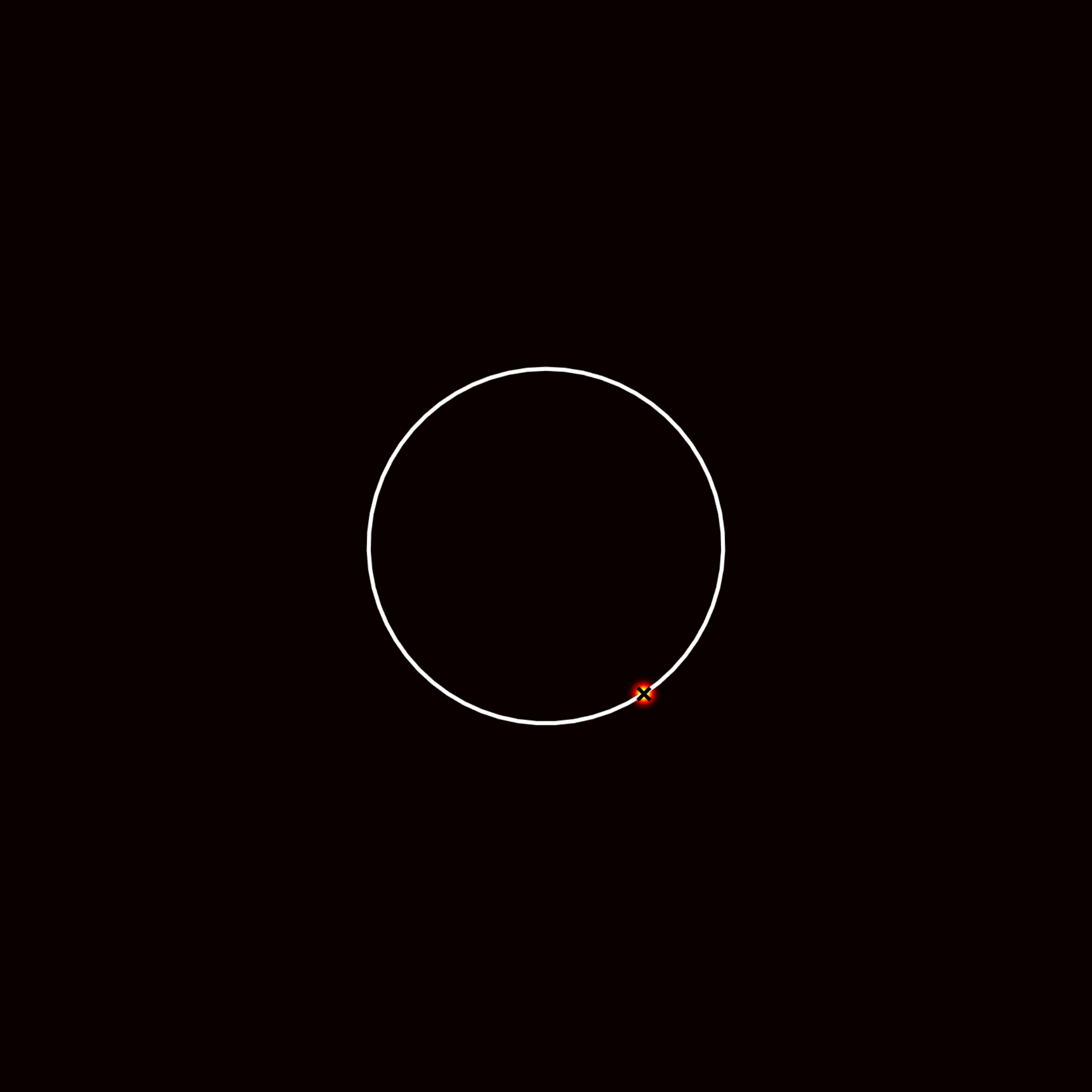}{0.48\textwidth}{(a)} \fig{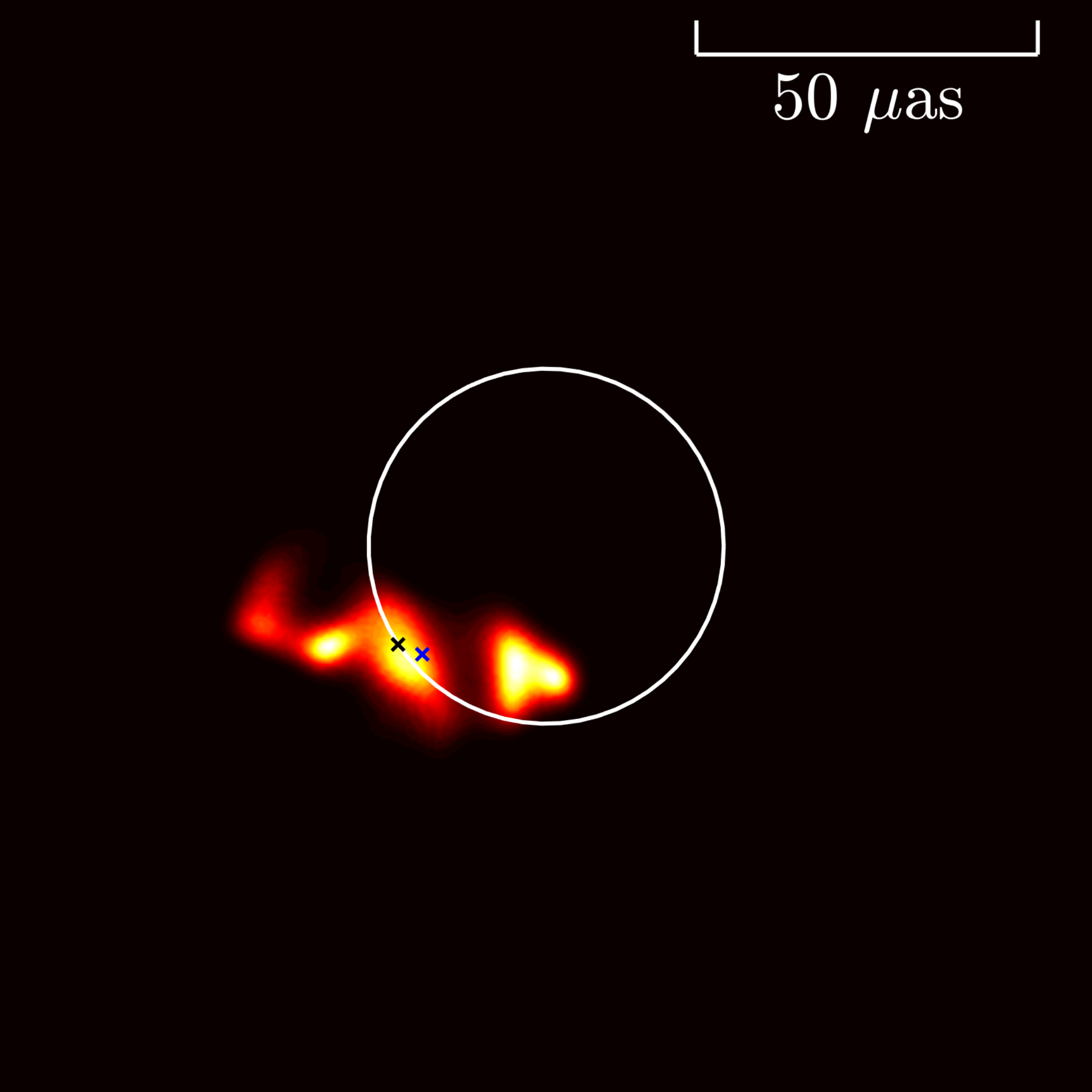}{0.48\textwidth}{(b)}}
    \caption{Example showing how the ring distortion is obtained. Panel (a) shows a Gaussian source with $\theta_\mathrm{src} = 2 \, \mu$as located on the ring at location (-21.7, -14.4)$ \, \mu$as with respect to the ring center. Panel (b) shows the corresponding scattered image using the GS06 model at $1.3\, \mathrm{mm}$. The white circle is a reference showing a uniform ring with $\theta_\mathrm{src} = 52 \, \mu$as. The black cross shows the image centroid of the unscattered image, and the blue cross shows the image centroid of the scattered Gaussian.}
\label{fig:gaussian_blobs}
\end{figure*}

Figure \ref{fig:ring_distortion} demonstrates visually the shape of the distorted ring using this method and compares the image distortion for two different scattering models. In both panels, the black ring shows the shape of the unscattered ring but is located at the image centroid of the scattered image. The dotted blue curves are obtained by tracing the relative image wander of 60 Gaussian sources located on the black circle. Figures~\ref{fig:ring_distortion}(a) and (b) show scattered images of a uniform ring using the J18 and GS06 models respectively. In panel (a), the distorted circle almost overlaps with the unscattered ring, whereas the ring is significantly distorted in panel (b). In addition, the features of the scattered image in panel (b) are smoother compared to those in panel (a) because the GS06 model has a very large inner scale (corresponding to an angular size of $5\,\mu{\rm as}$), which filters out high frequency components. As an example to demonstrate how the distorted ring with the GS06 model is obtained, Figure \ref{fig:gaussian_blobs} shows a scattered Gaussian with size $2\,\mathrm{\mu as}$ at one position on the ring. The flux in the scattered image is dispersed, which explains why the distorted ring in Figure \ref{fig:ring_distortion}(b) does not exactly match with the scattered image visually.

\setcounter{table}{0}
\begin{table}[h!]
\renewcommand{\thetable}{\arabic{table}}
\centering
\caption{Image wander and distortion of sources at 230 and 345 GHz with scattering models J18 and GS06. The image wander of a Gaussian source and the distortion for the J18 model are RMS fluctuation values obtained with the semi-analytic model; all other values are obtained with numerical simulations by taking the mean of 1500 realizations. The magnitude of distortion $|\Delta \bm \eta_\mathrm{0, a}|$ of each realization for the GS06 model is calculated by averaging over the distortion at 60 positions.} \label{tab:eta}
\begin{tabular}{ccccccccccc}
\tablewidth{0pt}
\hline
\hline
 & & \multicolumn5c{Image wander ($|\bm{\eta}_\mathrm{0, a}|$) ($\mu$as)} & \multicolumn2c{Distortion($|\bm{\Delta \eta}_\mathrm{0, a}|$) ($\mu$as)} & \\
\cline{3-10}
Model&  & \multicolumn2c{Gaussian} & \multicolumn2c{Ring} & \\ 
\cline{3-6}
 & & 230 GHz   &   345 GHz & 230 GHz   &   345 GHz & & 230 GHz  & 345 GHz\\
\hline
J18 &  &  &   & &  \\
$r_\mathrm{in} = 800\,\mathrm{km}$ &  & 0.532 &  0.271 & 0.714 & 0.478 & & 0.722 & 0.577 \\
 $\alpha = 1.38$  &  &  &   &  & \\
 \hline
GS06 &  &  &   &  &    \\
$r_\mathrm{in} = 2 \times 10^6\, \mathrm{km}$ &  & 2.09 &  1.35 & 3.82 &  2.92 & & 6.33 & 3.46\\
$\alpha = 0$  &  &  &  &  &  \\
\hline
\end{tabular}
\end{table}

An application of the framework we present is to provide an uncertainty for testing strong-field GR through the measurement of black hole shadow size. Since the magnitude of distortion decreases monotonically as a function of the source size (see Figure \ref{fig:deta_source_size}), calculating the relative image wander between two point sources (with $\theta_\mathrm{src} = 2 \, \mathrm{\mu as}$) separated by $26\,\mu$as provides an estimate for the maximum image distortion of Sgr~A* from scattering. Table \ref{tab:eta} summarizes the expected image wander and distortion for the two scattering models. Values of image wander are obtained from calculating the magnitude of image centroid shift of a Gaussian or circular source (ring). Except for the image wander of a ring, the data for the J18 model are the RMS fluctuation obtained with the semi-analytic framework using Equations (\ref{eqn:xrms}) and (\ref{eqn:detarms}). The image wander of a ring for the J18 model and all the data for the GS06 model are obtained with numerical simulations, and each value is calculated by taking the mean of 1500 realizations. The distortion $|\Delta \bm \eta_\mathrm{0, a}|$ for each realization is obtained by averaging over the centroid shift at 60 positions on the ring relative to a source at the center of the ring. We use numerical simulations for the GS06 model because when $\alpha = 0$, the linear approximation in Equation (\ref{eqn:i_approx}) is no longer valid, but the analytic framework relies on this approximation to compute the explicit form of $\fiw$ (see Equation (\ref{eqn:fiw})).

For the J18 model, the mean distortion from refractive scattering is $0.72\,\mathrm{\mu as}$, which is $1.4$\% of the source size and is significantly finer than the nominal angular resolution of near-term EHT observations. However, the GS06 model predicts a distortion that is an order of magnitude larger than the J18 model: the mean distortion from scattering is $6.3\,\mathrm{\mu as}$ ($\approx 12$\% of the source size), which can affect tests of GR. Note that the magnitudes of image wander and distortion are smaller at shorter wavelengths for both scattering models. For future submillimeter VLBI observations of Sgr~A* with a higher resolution, the image wander and distortion due to refractive scattering by the ISM will be less significant for both scattering models. However, even at 345 GHz, the mean distortion for the GS06 model is $3.46\,\mathrm{\mu as}$ ($\approx 6.7$\% of the source size), still larger than the $\sim$4\% effect from black hole spin and inclination for the Kerr metric.

\begin{figure*}[ht!]
\centering
\gridline{\fig{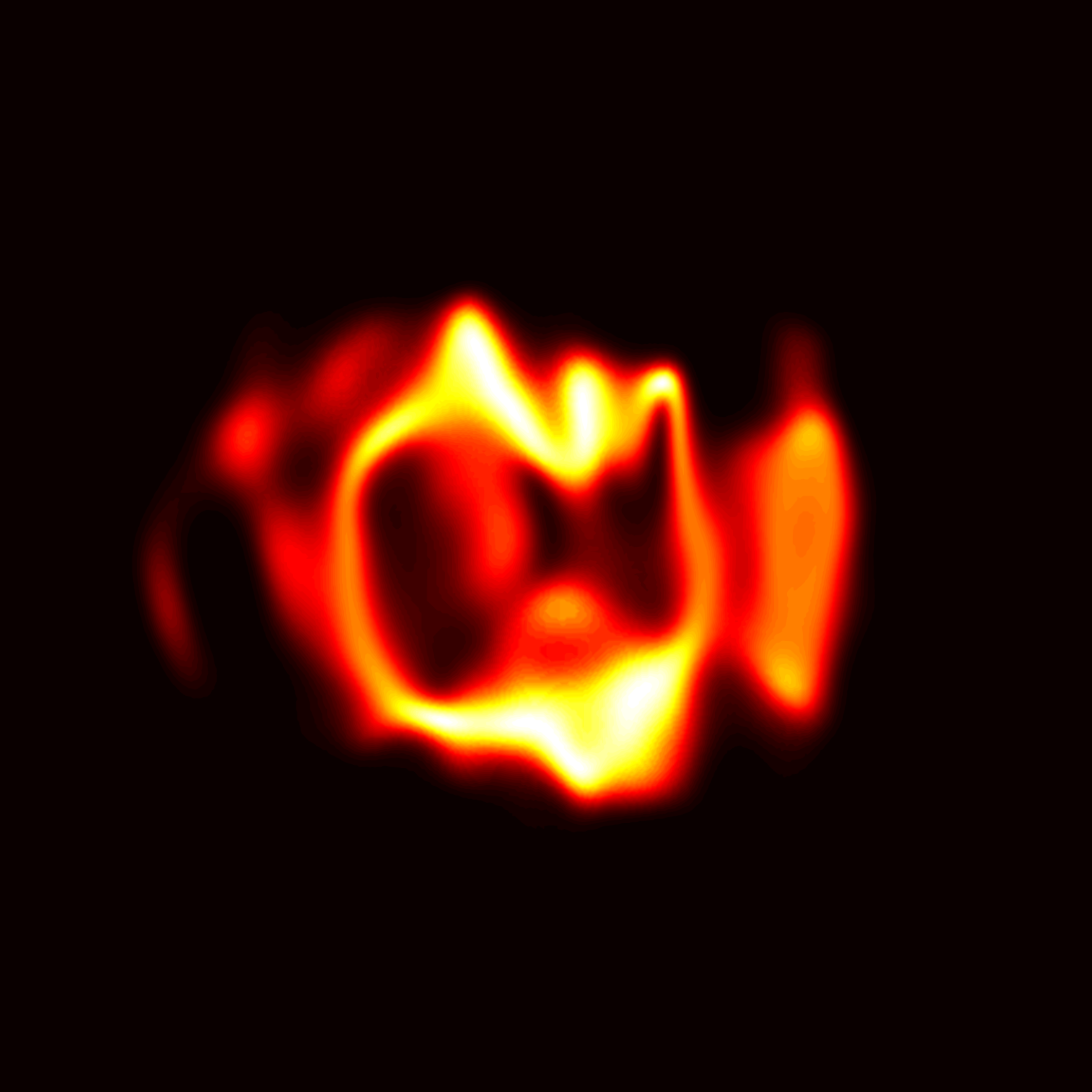}{0.24\textwidth}{(a)}\fig{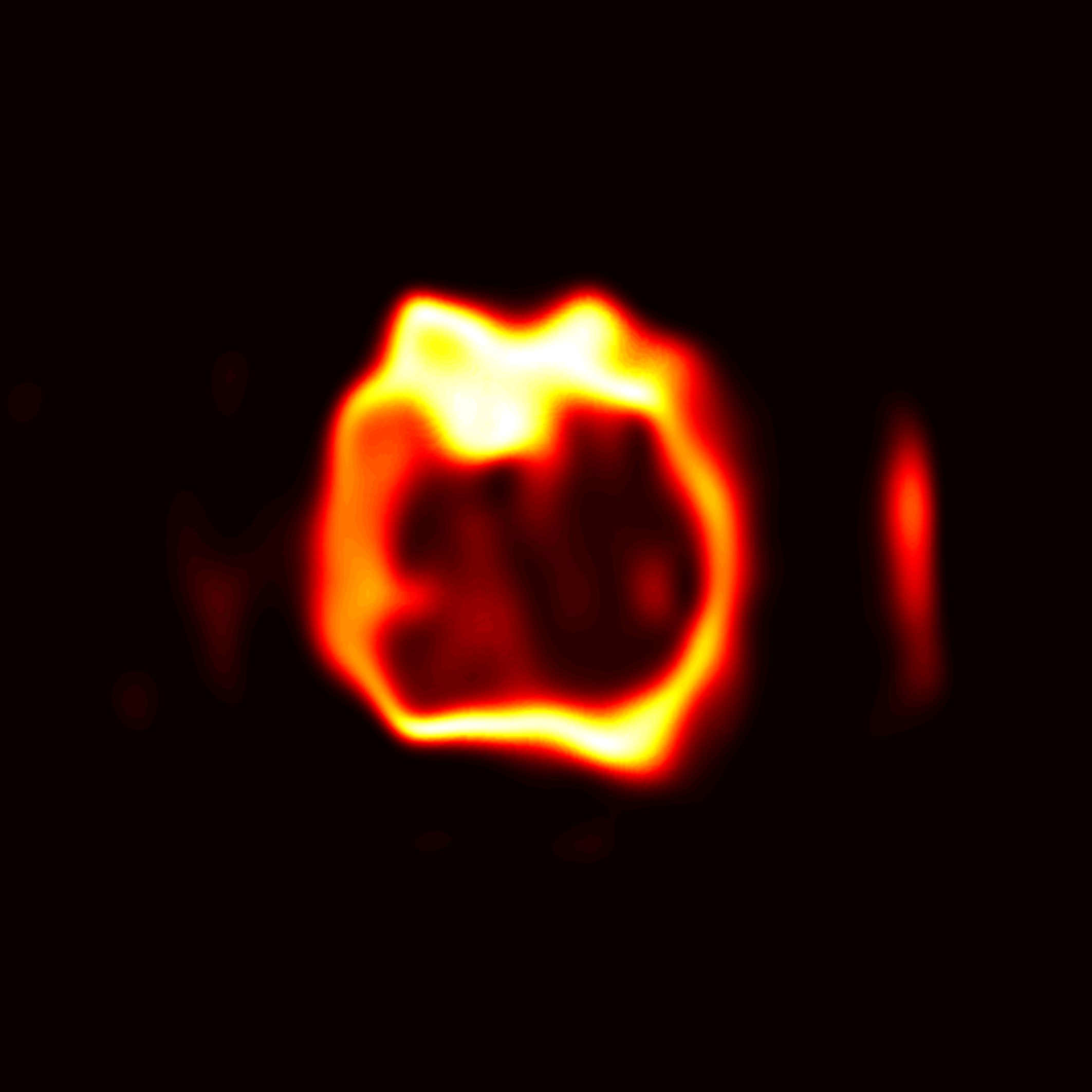}{0.24\textwidth}{(b)}\fig{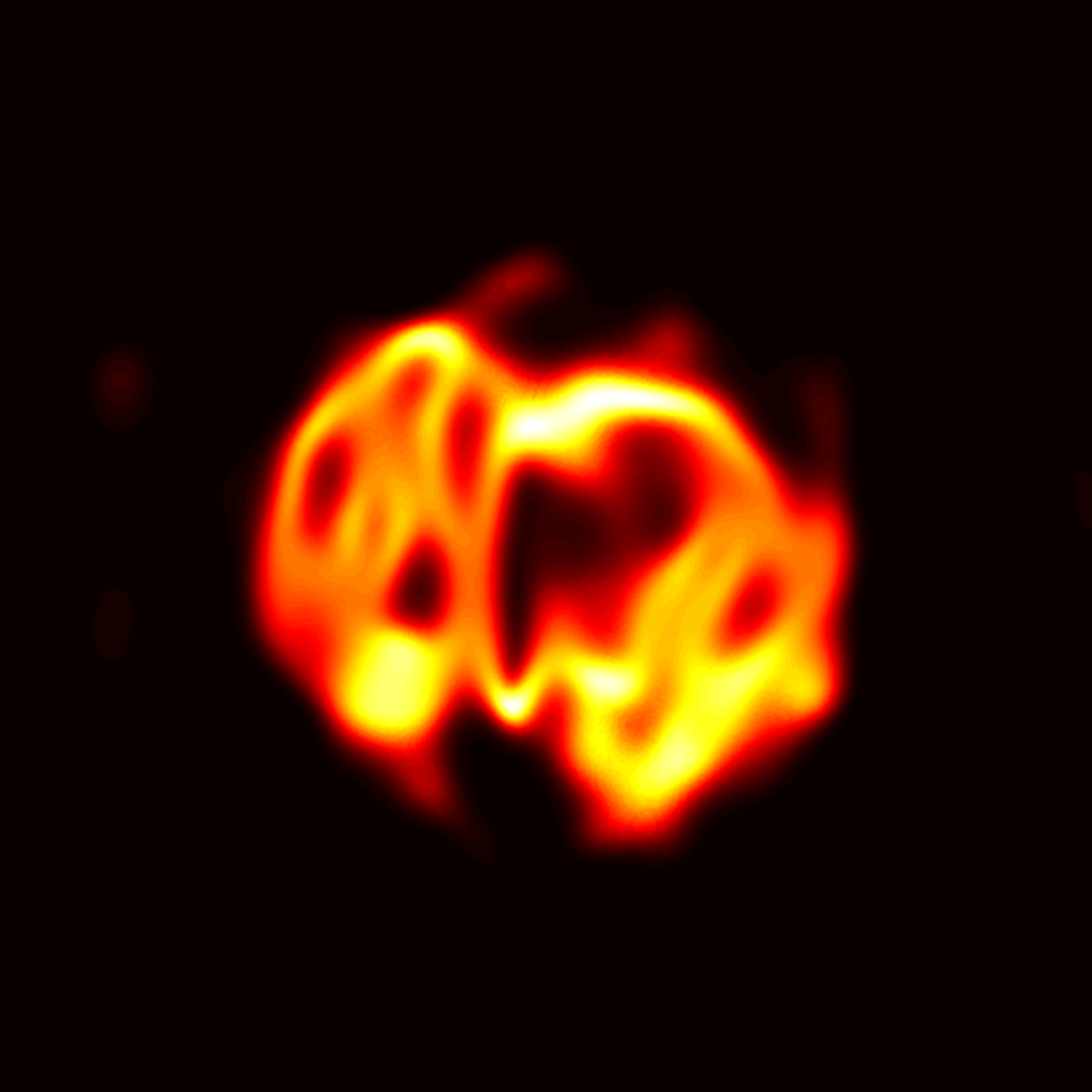}{0.24\textwidth}{(c)}\fig{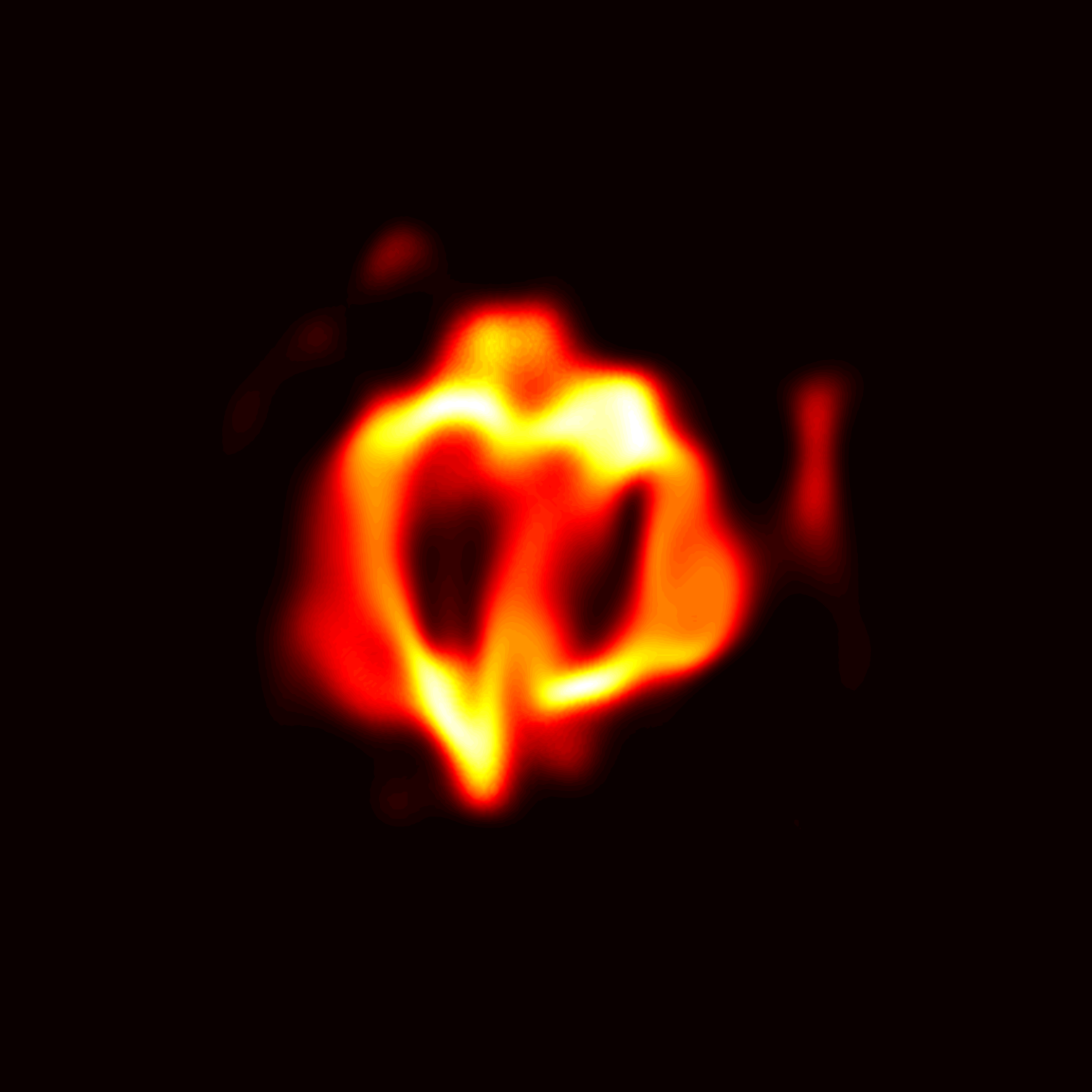}{0.24\textwidth}{(d)}}
    \caption{Examples showing a scattered ring with $\theta_\mathrm{src} = 52\,\mu$as using the GS06 model with $\alpha = 0$ and $r_\mathrm{\mathrm{in}} = 2 \times 10^6\, \mathrm{km}$ at $230\, \mathrm{GHz}$. The scattered images all show significant image distortion.}
    \label{fig:scattered_rings_model2}
\end{figure*}

In all cases, the GS06 model results in substantial image wander and image distortion. Figure \ref{fig:scattered_rings_model2} shows four realizations of a scattered ring using the GS06 model. Refractive scattering significantly distorts the original ring, which contrasts with the prediction in \citet{goldreich2006folded} that refractive scattering effects will be negligible in their model. This difference occurs because the required inner scale to match the observed refractive noise at 1.3\,cm is much larger than was originally proposed \citep{johnson2018scattering}. Moreover, this model was originally motivated by the assumption that the scattering screen is located near the source, which is now ruled out by measurements of temporal broadening of the Galactic Center magnetar \citep{Spitler_2014,Bower_Magnetar_2014,johnson2018scattering}. However, having $\alpha = 0$ and a large inner scale $r_\mathrm{in}$ does satisfy the constraints on the power index $\alpha$ and the inner scale $r_\mathrm{in}$ from measurements \citep[see Figure 14 in][]{johnson2018scattering}. Additional constraints from continued observations of Sgr~A* at millimeter wavelengths will allow for improved estimates of $\alpha$ and $r_{\rm in}$.

\subsection{Degree of Asymmetry of Sgr A*}\label{sec:asymmetry}
Another application of EHT images is to test the no-hair theorem using the degree of asymmetry $A$ (defined below) of the black hole shadow. For the Kerr metric, the asymmetry should be less than 0.6$\Theta_\mathrm{M}$ ($\sim 3\, \mu$as) for all inclination angles $i$ and spins $a$, where $\Theta_\mathrm{M} \equiv GM/Dc^2$ \citep[see Figure 7 in][]{chan2013gray}. Therefore, measuring an asymmetry $A> 3\, \mu$as in the shadow of Sgr~A* would indicate a violation of the no-hair theorem.  We now estimate the systematic uncertainty from scattering on measurements of $A$. 

In \citet{johannsen2010testing}, the asymmetry is defined for a closed continuous curve. Since distorted rings in our framework consist of finite numbers of points, we generalize the definition of $A$ by discretizing Equations (4)-(9) from \citet{johannsen2010testing}. First, define the center of the ring, $(\bar{x}, \bar{y})$, to be, 
\begin{align}
\bar{x} = \frac{1}{N} \sum^{N}_{i=1} x_i, \qquad \bar{y} = \frac{1}{N} \sum^{N}_{i=1} y_i,
\end{align}
where $(x_i, y_i)$ is the $i$-th position on the ring, and $N$ is the total number of positions. The displacement of the ring $D$ measures the shift of the center of the ring from the origin:
\begin{equation}
D = \sqrt{\bar{x}^2+\bar{y}^2}.
\end{equation}
The average radius of the ring is defined by the expression
\begin{equation}
\langle R \rangle \equiv \frac{1}{N} \sum_{i=1}^{N}R_i,
\label{eqn:ravg}
\end{equation}
where $R_i \equiv \sqrt{(x_i - \bar{x})^2 + (y_i-\bar{y})^2}$ is the distance from the $i$-th position to the center. We also define the ring diameter 
\begin{equation}
L = 2 \langle R \rangle.
\end{equation}
Finally, the degree of asymmetry is defined by the expression
\begin{equation}
A \equiv 2 \sqrt{\frac{\sum_{i=1}^{N}(R_i-\langle R \rangle)^2}{N}}.
\label{eqn:asymmetry}
\end{equation}
The degree of asymmetry $A$ can be computed for each distorted ring obtained with numerical simulations (i.e., the blue circles in Figure \ref{fig:ring_distortion}). 

\setcounter{table}{1}
\begin{table}[h!]
\renewcommand{\thetable}{\arabic{table}}
\centering
\caption{Asymmetry and shift in diameter of sources at 230 and $345\, \mathrm{GHz}$ with scattering models J18 and GS06. Each value, along with the 95\% range, is obtained from 1500 realizations. } \label{tab:asym}
\begin{tabular}{cccccc}
\tablewidth{0pt}
\hline
\hline
 & & \multicolumn{2}{c}{Asymmetry ($A$) $\mathrm{(\mu as)}$} & \multicolumn{2}{c}{Diameter Shift ($\Delta L$) $\mathrm{(\mu as)}$} \\
\cline{2-6}
& &  230 GHz  & 345 GHz & 230 GHz & 345 GHz \\
\hline
J18 & { } & { } & { }  \\
$r_\mathrm{in} = 800\,\mathrm{km}$ &  & $0.515^{+0.289}_{-0.195}$ &  $0.361^{+0.177}_{-0.119}$ &  $0.006^{+0.925}_{-0.897}$&  $0.010^{+0.571}_{-0.564}$ \\
 $\alpha = 1.38$  & { } & { } & { }  & { } \\
\hline
GS06 & { } & { } &  { } & { } &    \\
$r_\mathrm{in} = 2 \times 10^6\, \mathrm{km}$ &  & $5.01^{+3.13}_{-2.22}$  & $2.76^{+1.46}_{-1.18}$ & $0.27^{+5.43}_{-5.10}$  & $0.17^{+3.95}_{-3.84}$  \\
$\alpha = 0$ & & & &  \\
\hline
\end{tabular}
\end{table}

\begin{figure*}[ht!]
\centering
\gridline{\fig{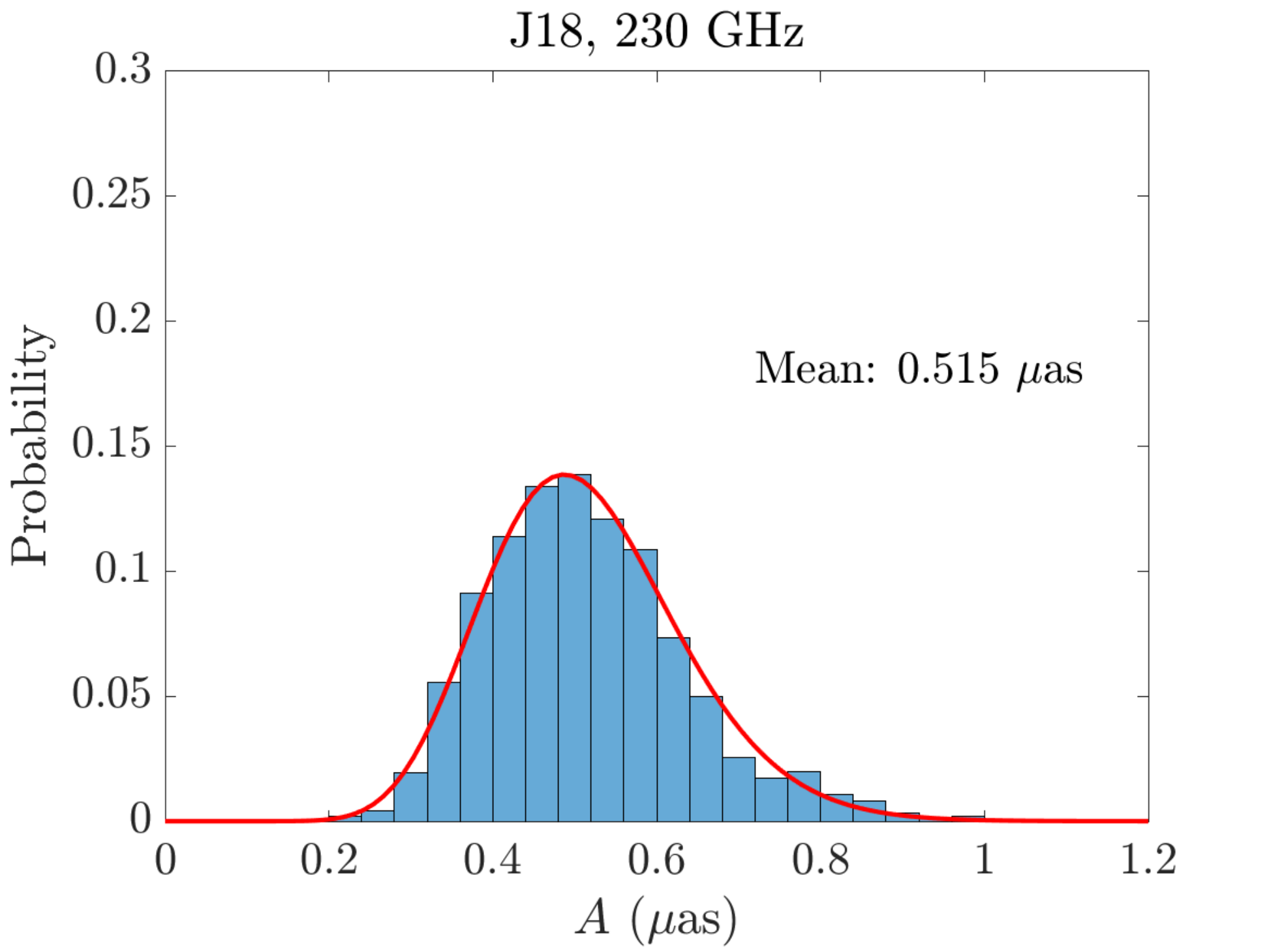}{0.48\textwidth}{(a)}
         \fig{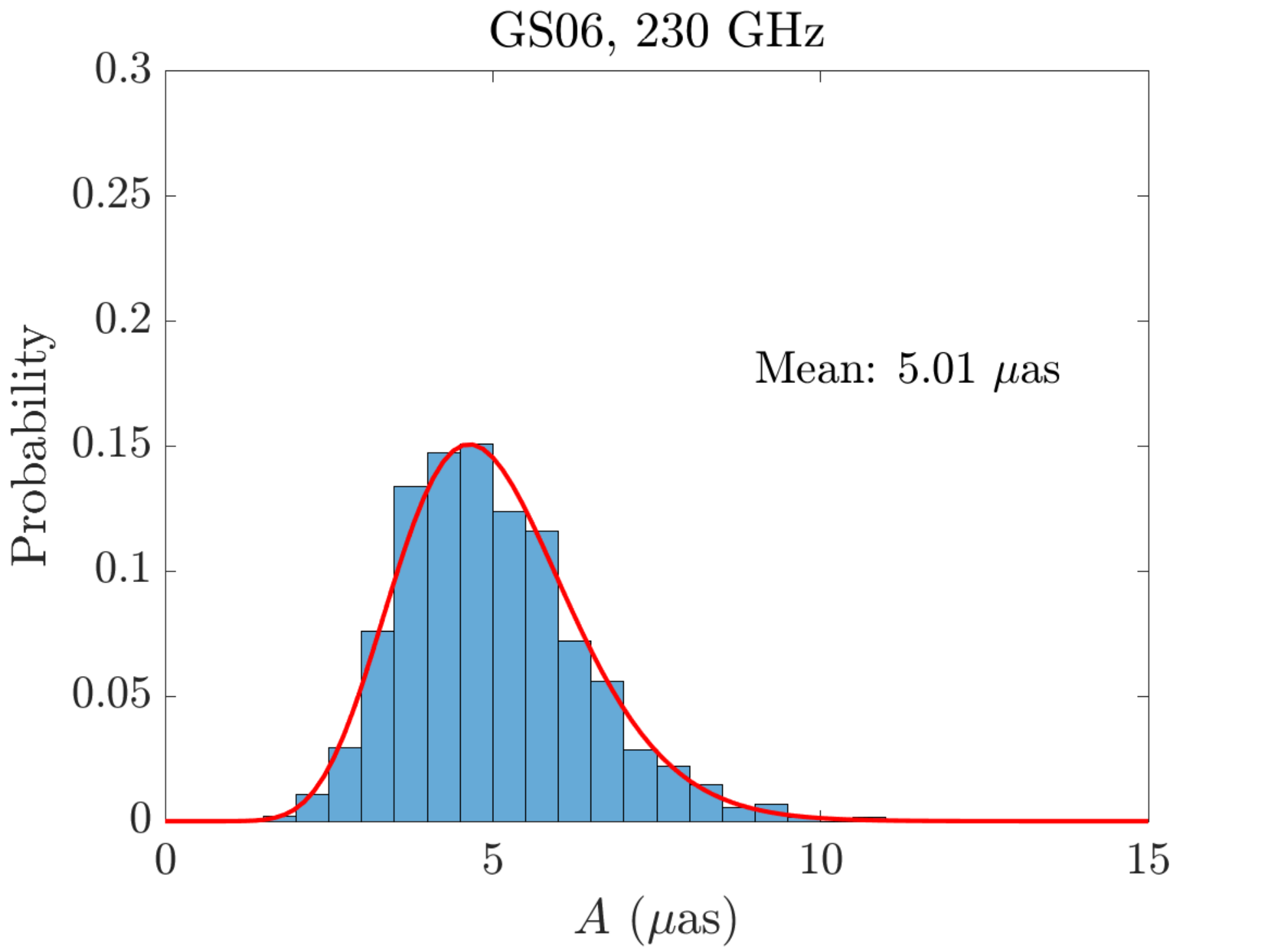}{0.48\textwidth}{(b)}}
        {\fig{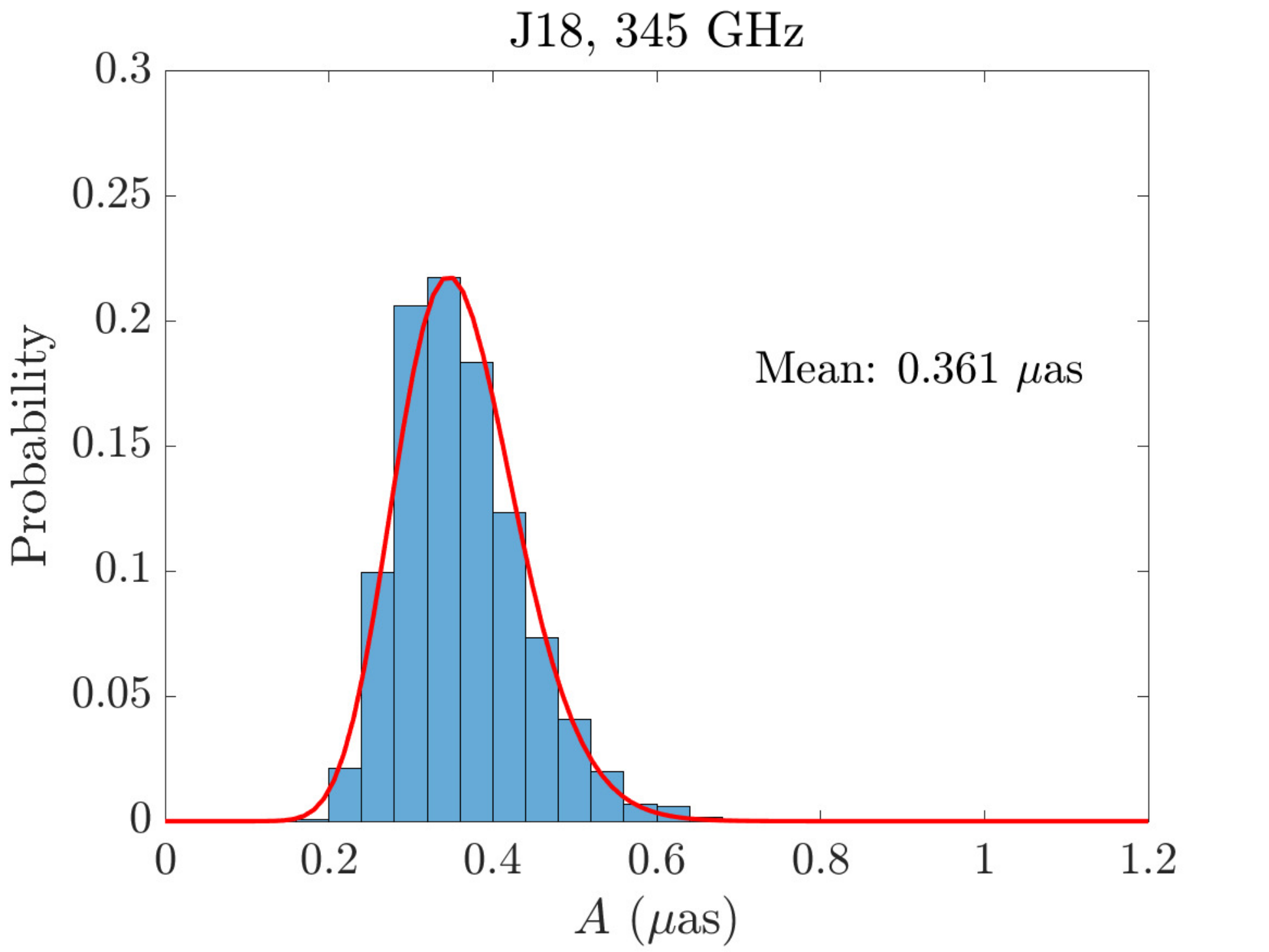}{0.48\textwidth}{(c)}
         \fig{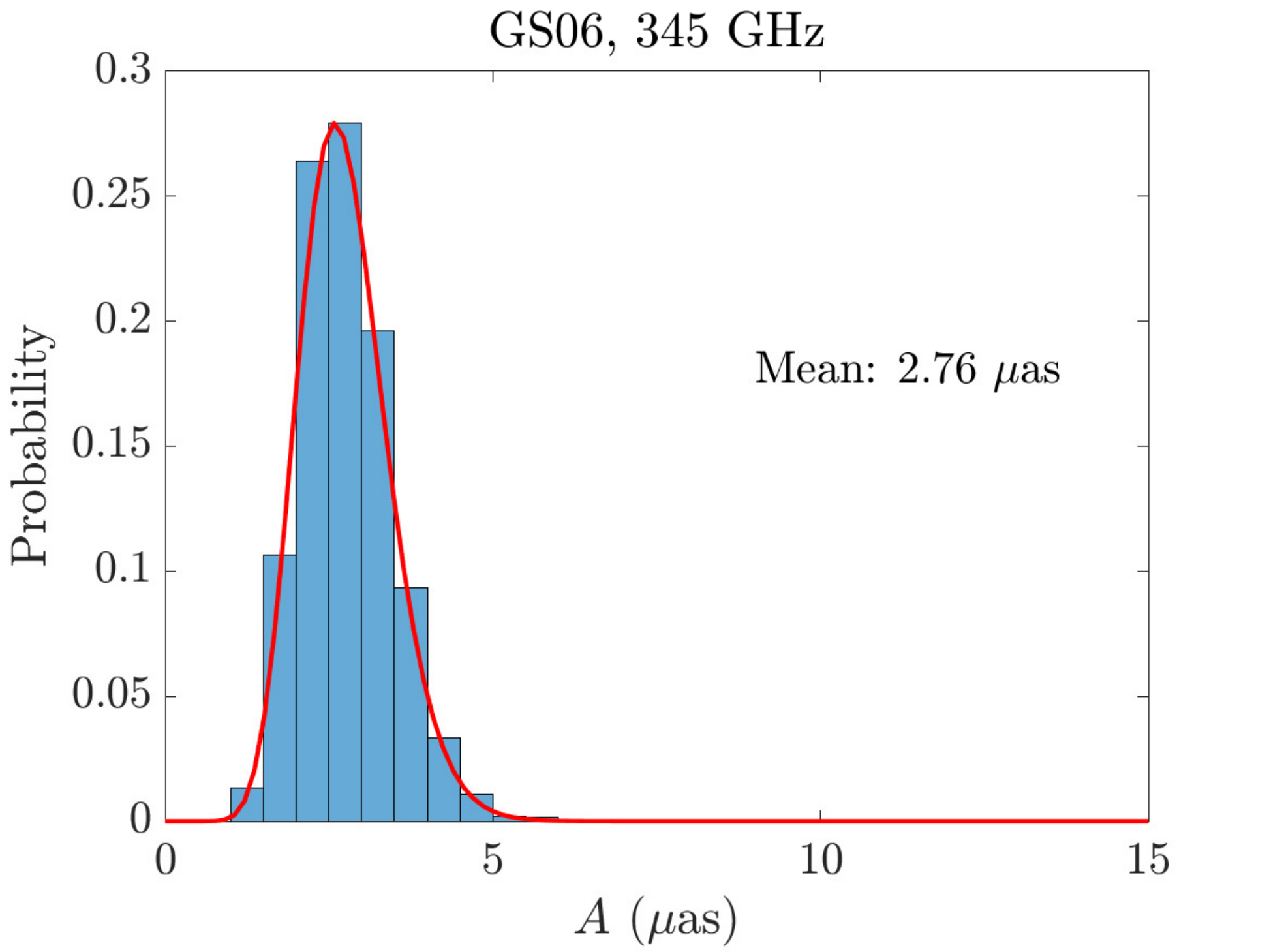}{0.48\textwidth}{(d)}}
    \caption{Distributions of the scattering-induced degree of asymmetry $A$ for the J18 and GS06 models at 230 and $345\, \mathrm{GHz}$, which exhibit tails extending to large asymmetries. The red curves are fits to a Gamma distribution. }
    \label{fig:A_dist}
\end{figure*}

Table \ref{tab:asym} summarizes the mean values of the asymmetry and the shift in diameter, as well as their 95\% ranges. These estimates are calculated using 1500 independent scattering realizations with 60 positions on the ring for each realization. Using the J18 model, the mean asymmetry at both $230$ and $345\, \mathrm{GHz}$ is within the $0.6\Theta_\mathrm{M}$ range spanned by varying black hole spin and inclination, whereas the asymmetry from the GS06 model has the same order of magnitude as this $0.6\Theta_\mathrm{M}$ range. The distributions of asymmetry can be well fitted to a Gamma distribution (the red curves in Figure \ref{fig:A_dist}),
\begin{equation}
y = f(x|a, b) = \frac{1}{b^a \Gamma(a)} x^{a-1} \e^{-\frac{x}{b}},
\end{equation}
where $\Gamma(a) = \int_0^{\infty} x^{a-1} \e^{-x} \mathrm{dx}$ is the Gamma function. The distributions of $A$ have long tails into large values; i.e., scattering will occasionally contribute anomalously large systematic errors for tests of the no-hair theorem. The mean shift of diameter is positive in all four cases, which implies that scattering preferentially stretches the ring by a small amount.

\subsection{Uncertainty Reduction in the Ensemble Average Regime}

\setcounter{table}{2}
\begin{table}[h!]
\renewcommand{\thetable}{\arabic{table}}
\centering
\caption{The change in diameter $\Delta L$ and degree of asymmetry $A$ of distorted rings with $\theta_\mathrm{src} = 52\, \mu$as of $n$ averaged images at $1.3\, \mathrm{mm}$ using the J18 model. Each value, along with its 95\% range, is calculated by taking the mean of $n$-averaged images.} \label{tab:asymmetry_avg_img}
\begin{tabular}{ccccccccc}
\tablewidth{0pt}
\hline
\hline
$n$ & 1 & 2 & 4 & 5 & 10 & 20 \\
 \hline
$\Delta L$ ($\mu$as) & $0.006^{+0.925}_{-0.897}$ & $0.005^{0.662}_{-0.650}$ &$0.004^{+0.467}_{-0.471}$ & $0.004^{+0.419}_{-0.410}$ & $0.004^{+0.257}_{-0.322}$ & $0.004^{+0.188}_{-0.246}$ \\
 \hline
$A$ ($\mu$as) & $0.515^{+0.289}_{-0.195}$ & $0.360^{0.160}_{-0.141}$ & $0.254^{0.133}_{-0.097}$ & $0.228^{+0.104}_{-0.083}$ & $0.163^{+0.071}_{-0.062}$& $0.114^{+0.056}_{-0.045}$ \\
\hline
\end{tabular}
\end{table}
The ring radius of an average image from multiple observations may be determined using feature extraction techniques \citep[e.g.,][]{nixon2012feature,psaltis2015general,Kuramochi_2018}, and the corresponding asymmetry can be computed. One way to reduce asymmetry from scattering is to average the scattered image over a number of scattering realizations. In practice, averaging over realizations is equivalent to performing a temporal average, and the coherence timescale of the scattering at 230\,GHz is expected to be approximately 1 day \citep{johnson2016stochastic}. As the number of averages approaches infinity, the average image approaches the ensemble-average regime. In this regime, the only effect from scattering is diffractive blurring, which has no associated centroid shift. Thus, in the ensemble-average regime, any residual image asymmetry (as defined above) will be intrinsic. Table \ref{tab:asymmetry_avg_img} summarizes mean values and 95\% ranges of $\Delta L$ and $A$ of $n$-averaged scattered images. To compute these values, we used the same ensemble of 1500 scattered images as in Section \ref{sec:asymmetry}. As $n$ increases, the distribution narrows, with the mean of $A$ and $\Delta L$ being proportional to $1/\sqrt{n}$. Using images from a single EHT campaign, which typically lasts for five to six days, the estimated uncertainty of $A$ from scattering could be reduced by more than a factor of 2 compared to that without image averaging. However, if the coherence timescale is longer than 1 day (e.g., from a lower than expected effective velocity or at longer wavelengths), then corresponding longer campaigns would be required to reduce the uncertainty from scattering. 

\section{Summary}\label{sec:conclusion}
With rapidly increasing angular resolution and sensitivity, VLBI observations are now capable of producing images at the event-horizon scale. Forthcoming EHT images of Sgr~A* can potentially measure its shadow size and shape, enabling a new test of strong-field GR. However, with advances in resolution and sensitivity, substructure due to scattering has also become observable \citep[see e.g.,][]{gwinn2014discovery, ortiz2016intrinsic, johnson2018scattering} and distorts the image of Sgr~A*. Therefore, it is essential to understand how refractive scattering can affect measurements of the shadow size and asymmetry.

In this paper, we derived a framework to quantify image wander and distortion from scattering. The results depend on the unscattered image, the ensemble-average scatter-broadening kernel, and the power spectrum of density fluctuations within the scattering material. 
We showed that the currently favored scattering model for Sgr~A*, J18, does not substantially affect the shadow size and shape, which implies that refractive effects do not impair the ability of the EHT to test GR. We estimate the mean refractive image wander, distortion, and asymmetry to be 0.53$\, \mu$as, 0.72$\, \mu$as, and 0.52$\, \mu$as at $230\, \mathrm{GHz}$, and $0.27\, \mu$as, $0.58\, \mu$as, and $0.36\, \mu$as at $345 \, \mathrm{GHz}$. However, an alternative scattering model, GS06, has a flatter power spectrum and requires a large inner scale in order to fit observations of Sgr~A*, causing significant image distortion at millimeter wavelengths. For this model, we estimate the mean image wander, distortion and asymmetry to be $2.1\, \mu$as, $6.3\, \mu$as, and $5.0\,\mu$as at $230\, \mathrm{GHz}$, and $1.4\, \mu$as, $3.5\, \mu$as, and $2.8\, \mu$as at $345 \, \mathrm{GHz}$. In both cases, we showed that taking the average image from multiple observations can reduce these effects from scattering. 


In short, while the significant differences between the J18 and GS06 models demonstrate the necessity for tighter constraints on the refractive scattering properties of Sgr~A* at millimeter wavelengths, uncertainties from refractive scattering are unlikely to dominate the error budget for EHT measurements. 



\acknowledgements{We thank the National Science Foundation (AST-1716536, OISE-1743747) and the Gordon and Betty Moore Foundation (GBMF-5278) for financial support. This work was also supported in part by the Black Hole Initiative at Harvard University, which is supported by a grant from the John Templeton Foundation. Z.Z was supported by the Purcell fellowship and the Peirce fellowship.}

\appendix
\section{Detailed Calculation of Image Wander for a Gaussian Source with an Isotropic Power Spectrum}
Equation (\ref{eqn:xrms}) provides a closed form expression for the RMS refractive fluctuations of image wander given the source image and the power spectrum. Here, we present a detailed calculation of image wander using the example of a Gaussian source and an isotropic power spectrum for the turbulent ISM.

The structure function $D_\phi (\bm r)$ of an isotropic power spectrum with power index $\alpha$ takes the form $D_\phi (\bm r) = \powerlaw$. Using Equation (\ref{eqn:Qq_of_Dphi}), we can obtain the power spectrum $Q(\bq)$ at $\bq \neq \bm{0}$,
\begin{align}
Q(\bm q) & = 2^\alpha \pi \alpha \frac{\Gamma (1 + \frac{\alpha}{2})}{\Gamma (1 - \frac{\alpha}{2})} \lambdabar^{-2} \bm{r}_\mathrm{diff}^{-\alpha} \left |\bm q \right|^{-(2 + \alpha)}.
\label{eqn:spectrum}
\end{align}

The intensity of a Gaussian source takes the  form $I_\mathrm{\mathrm{src}} (\bm r) = I_\mathrm{\mathrm{src}} (\bm 0) \e^{-4 \ln 2\bm (r^2 / w^2)}$, where $w$ is the full width at half maximum of the Gaussian: $w = \theta_\mathrm{src} D$. In the Fourier domain, the visibility of the Gaussian source is,
\begin{align}
V_\mathrm{\mathrm{src}} (\bm b) & = V_\mathrm{\mathrm{src}} (\bm 0) \e^{-\frac{1}{2} \left(\frac{\theta_\mathrm{src} \bb}{2\sqrt{2 \ln 2} \lambdabar}\right)^2}
\label{eqn:v_scr}
\end{align}
The ensemble average visibility is obtained by multiplying the visibility of the source by the scattering kernel in the Fourier space $\tilde{G}(\bm b)$ given in Equation (\ref{eqn:kernel_ft}),
$V_\mathrm{ea} (\bb) = V_\mathrm{\mathrm{src}} (\bm b) \e^{-\frac{1}{2} D_\phi \left(\frac{\bb}{1+M}\right)}$. To simplify the calculation, here we approximate $D_\phi(\bm r) \approx |\frac{\bm r}{r_\mathrm{diff}}|^2$ such that the ensemble average visibility is still a Gaussian,
\begin{align}
V_\mathrm{ea} (\bb)& = V_\mathrm{ea} (\bm 0) \exp \left[ - \frac{1}{2} \left(\frac{\theta_\mathrm{ea}}{\theta_\mathrm{scatt}}\frac{\bb}{(1+M) r_\mathrm{\mathrm{diff}}}\right)^2 \right],
\end{align} 
where $\theta_\mathrm{ea} = \sqrt{\theta_\mathrm{src}^2 + \theta_\mathrm{scatt}^2}$.

We then evaluate $\fiwq$ for the Gaussian source by taking the gradient of $\fvq$ at the zero baseline (Equation (\ref{eqn:fiwq})). We first obtain $\fvq$ by taking the Fourier transform of $\fv$, which is given in Equation (\ref{eqn:fv}),  
\begin{align}
\fvq & = r_\mathrm{\mathrm{F}}^2 \intr \ \fv \ft \nonumber \\
& = r_\mathrm{\mathrm{F}}^2 \bq \cdot \left(\frac{\bb}{D \lambdabar} + \bm q \right) V_\mathrm{ea}(D \lambdabar \bm q + \bm b).
\end{align}
From this, we calculate $\fiwq$ that maps to refractive image wander as follows, 
\begin{align}
\fiwq & =  \frac{i \lambdabar}{V_\mathrm{ea} (\bm 0)} \partialb \fvq \nonumber \\
& = \frac{i\lambdabar}{V_\mathrm{ea} (\bm 0)} r_\mathrm{F}^2 \left[\qdl V_\mathrm{ea}(D \lambda \bm q + \bm b) + \bm q \left (\qdl + \bm q \right) \grad_\mathrm{\bb} V_\mathrm{ea}(D \lambdabar \bm q + \bm b)  \right] |_\mathrm{\bm b = \bm 0} \nonumber \\
& =\frac{i \lambdabar }{M+1} \left(\bq-\frac{\theta_\mathrm{ea}}{\theta_\mathrm{diff}} r_\mathrm{ref}^2 \bq^3\right) \exp \left[ - \frac{1}{2} \left( \frac{\theta_\mathrm{ea}}{\theta_\mathrm{diff}} r_\mathrm{ref} \bq\right)^2 \right],
\label{eqn:fq}
\end{align}
where $r_\mathrm{ref} = \frac{r_\mathrm{\mathrm{F}}^2}{r_\mathrm{\mathrm{diff}}}$, and $D\lambdabar = r_\mathrm{\mathrm{F}}^2 (1 + M)$. 

\begin{figure*}[ht]
\gridline{\fig{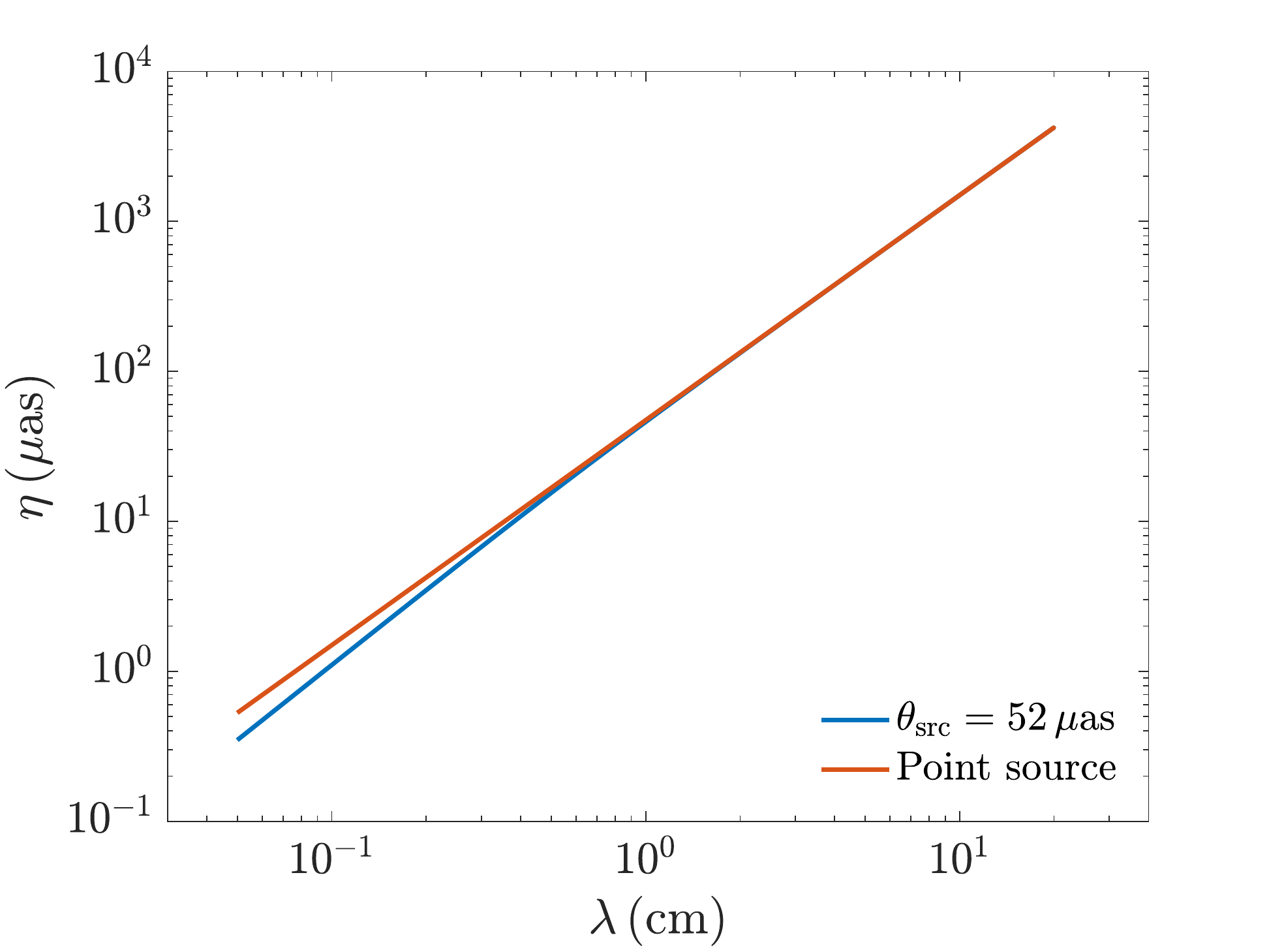}{0.49\textwidth}{(a)}
          \fig{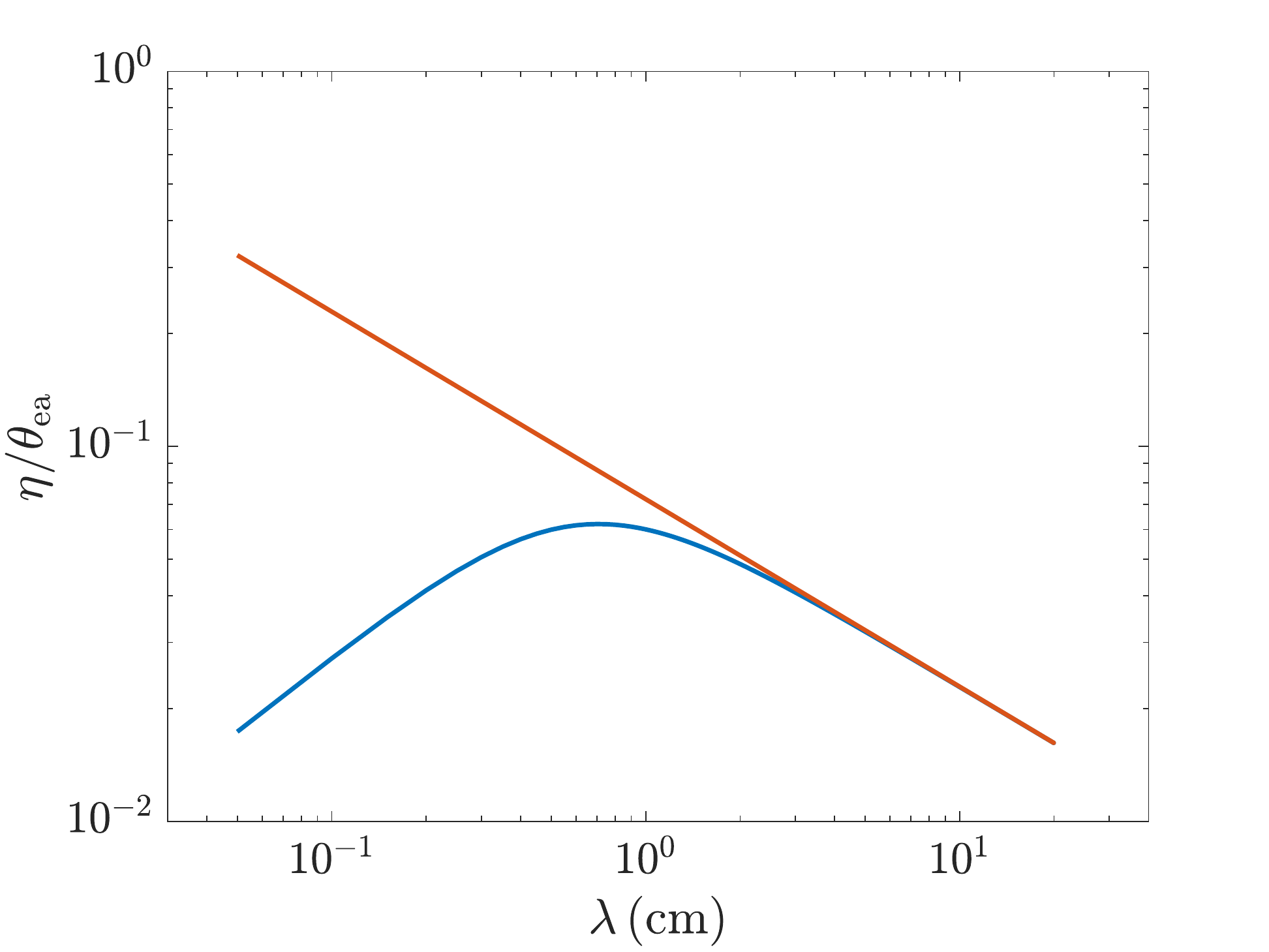}{0.49\textwidth}{(b)}
          }
\caption{Panel (a) and (b) show RMS and fractional position wander as a function of observing wavelength respectively, which are obtained from the closed-form analytic expression given in Equation (\ref{eqn:etarms_iso}). Curves shown use source and scattering parameters similar to those of Sgr~A*: $\theta_\mathrm{src} = 400 \lambda_\mathrm{cm} \, \mathrm{\mu as}$, $r_\mathrm{diff} = 410\, \mathrm{km} / \lambda_\mathrm{cm}$, $r_\mathrm{F} = \sqrt{DR\lambdabar/(D + R) }$, $D = 2.8 \, \mathrm{kpc}$, $ R = 5.3 \, \mathrm{kpc}$, and $\alpha = 5/3$.} \label{fig:eta_vs_lambda}
\end{figure*}

With $Q(\bq)$ and $\fiwq$, we can now obtain an analytic expression for mean-squared refractive image wander using Equation (\ref{eqn:xrms}), 
\begin{align}
\etarms & = \left(\frac{\lambdabar}{2\pi(1+M)}\right)^2 2^\alpha \pi \alpha \frac{\Gamma \left(1+\frac{\alpha}{2}\right)}{\Gamma \left(1-\frac{\alpha}{2}\right)} \intq\ |\bq|^{-(2 + \alpha)} \left[ \bq-\left(\frac{\theta_\mathrm{ea}}{\theta_\mathrm{scatt}}r_\mathrm{ref}\right)^2 \bq^3 \right]^2 \nonumber \\
& \times \exp \left[ - \frac{1}{2} \left(\frac{\theta_\mathrm{ea}}{\theta_\mathrm{scatt}} r_\mathrm{ref} \bq\right)^2 \right].
\label{eqn:xrms_int}
\end{align}
We then evaluate the integral analytically in polar coordinates by letting $|\bq| = \sqrt{q_x^2 + q_y^2}$, where $q_x = q \cos \theta, q_y = q \sin \theta$. 
\begin{align}
\etarms & = \left(\frac{\lambdabar}{1+M}\right)^2 2^{\alpha-1} \alpha \frac{\Gamma \left(1+\frac{\alpha}{2}\right)}{\Gamma \left(1-\frac{\alpha}{2}\right)} \int_0^\infty q_r^{-(2 + \alpha)} \left[ q_r-\left(\frac{\theta_\mathrm{ea}}{\theta_\mathrm{scatt}}r_\mathrm{ref}\right)^2 q_r^3 \right]^2 \exp \left[ - \frac{1}{2} \left(\frac{\theta_\mathrm{ea}}{\theta_\mathrm{scatt}} r_\mathrm{ref} q_r\right)^2 \right] \nonumber \\
& = \frac{2^{\alpha -6} \Gamma \left(\frac{\alpha }{2}+1\right) \alpha  (\alpha^2-2\alpha+4) }{\pi ^2 (M+1)^2} \left(\frac{r_\mathrm{\mathrm{diff}}}{r_\mathrm{F}}\right)^{4-2\alpha} \left(\frac{\theta_\mathrm{scatt}}{\theta_\mathrm{F}}\right)^{2-\alpha} \left(\frac{\lambda}{r_\mathrm{\mathrm{diff}}} \right)^2.
\end{align}
Finally, the RMS refractive angular fluctuation of the image wander is, 
\begin{align}
\sqrt{\etarms} & = \frac{2^{\alpha/2 -3}}{\pi (M+1)} \sqrt{\Gamma \left(\frac{\alpha }{2}+1\right) \alpha  (\alpha^2-2\alpha+4)} \left(\frac{r_\mathrm{\mathrm{diff}}}{r_\mathrm{F}}\right)^{2-\alpha} \left(\frac{\theta_\mathrm{scatt}}{\theta_\mathrm{ea}}\right)^{1-\alpha/2} \left(\frac{\lambda}{r_\mathrm{\mathrm{diff}}} \right).
\label{eqn:etarms_iso}
\end{align} 

Equation (\ref{eqn:etarms_iso}) suggests that the RMS position wander scales as $\left(\frac{r_\mathrm{\mathrm{diff}}}{r_\mathrm{F}}\right)^{2-\alpha} \left(\frac{\theta_\mathrm{scatt}}{\theta_\mathrm{ea}}\right)^{1-\alpha/2} \left(\frac{\lambda}{r_\mathrm{\mathrm{diff}}} \right)$, which matches with the results of \cite{cordes1986refractive}, \cite{romani1986refractive}, and \cite{johnson2015theory}. For a point source, the position wander scales as $\lambda^{3/2\alpha - 1}.$ The position wander can also be normalized by the angular size of the ensemble average image. The normalized image wander scales as  $\left(\frac{r_\mathrm{\mathrm{diff}}}{r_\mathrm{F}}\right)^{2-\alpha} \left(\frac{\theta_\mathrm{scatt}}{\theta_\mathrm{ea}}\right)^{2-\alpha/2} $. Figure \ref{fig:eta_vs_lambda} shows the RMS and fractional refractive image wander as a function of the observing wavelength for an isotropic Kolmogorov ($\alpha = 5/3$) spectrum. Note that at this power law index, the magnitude of refractive image wander increases as a function of $\lambda$, whereas most other refractive features tend to weaken. 

\bibliographystyle{aasjournal}
\bibliography{mybib}

\begin{thebibliography}{}
\expandafter\ifx\csname natexlab\endcsname\relax\def\natexlab#1{#1}\fi
\providecommand{\url}[1]{\href{#1}{#1}}

\bibitem[{Bardeen(1973)}]{bardeen1973black}
Bardeen, J. 1973, DeWitt and BS DeWitt (New York: Gordon and Breach), 215

\bibitem[{Blandford \& Narayan(1985)}]{blandford1985low}
Blandford, R., \& Narayan, R. 1985, \mnras, 213, 591

\bibitem[{Boehle {et~al.}(2016)Boehle, Ghez, Sch{\"o}del, Meyer, Yelda, Albers,
  Martinez, Becklin, Do, Lu, {et~al.}}]{boehle2016improved}
Boehle, A., Ghez, A., Sch{\"o}del, R., {et~al.} 2016, \apj, 830, 17

\bibitem[{{Bouman} {et~al.}(2017){Bouman}, {Johnson}, {Dalca}, {Chael},
  {Roelofs}, {Doeleman}, \& {Freeman}}]{Bouman_2017}
{Bouman}, K.~L., {Johnson}, M.~D., {Dalca}, A.~V., {et~al.} 2017, Transactions
  on Computational Imaging, submitted (arXiv:1711.01357)

\bibitem[{{Bower} {et~al.}(2014){Bower}, {Deller}, {Demorest}, {Brunthaler},
  {Eatough}, {Falcke}, {Kramer}, {Lee}, \& {Spitler}}]{Bower_Magnetar_2014}
{Bower}, G.~C., {Deller}, A., {Demorest}, P., {et~al.} 2014, \apjl, 780, L2

\bibitem[{Bower {et~al.}(2015)Bower, Deller, Demorest, Brunthaler, Falcke,
  Moscibrodzka, O'Leary, Eatough, Kramer, Lee, {et~al.}}]{bower2015proper}
Bower, G.~C., Deller, A., Demorest, P., {et~al.} 2015, \apj, 798, 120

\bibitem[{Carter(1971)}]{carter1971axisymmetric}
Carter, B. 1971, \prl, 26, 331

\bibitem[{Carter(1973)}]{carter1973black}
---. 1973, Black holes, 57

\bibitem[{Chael {et~al.}(2018)Chael, Rowan, Narayan, Johnson, \&
  Sironi}]{chael2018role}
Chael, A., Rowan, M., Narayan, R., Johnson, M., \& Sironi, L. 2018, \mnras

\bibitem[{{Chael} {et~al.}(2018){Chael}, {Johnson}, {Bouman}, {Blackburn},
  {Akiyama}, \& {Narayan}}]{Chael_Closure}
{Chael}, A.~A., {Johnson}, M.~D., {Bouman}, K.~L., {et~al.} 2018, \apj, 857, 23

\bibitem[{Chael {et~al.}(2016)Chael, Johnson, Narayan, Doeleman, Wardle, \&
  Bouman}]{chael2016high}
Chael, A.~A., Johnson, M.~D., Narayan, R., {et~al.} 2016, \apj, 829, 11

\bibitem[{Chan {et~al.}(2013)Chan, Psaltis, \& {\"O}zel}]{chan2013gray}
Chan, C.-k., Psaltis, D., \& {\"O}zel, F. 2013, \apj, 777, 13

\bibitem[{Coles {et~al.}(1987)Coles, Rickett, Codona, \&
  Frehlich}]{coles1987refractive}
Coles, W.~A., Rickett, B., Codona, J., \& Frehlich, R. 1987, \apj, 315, 666

\bibitem[{Cordes {et~al.}(1986)Cordes, Pidwerbetsky, \&
  Lovelace}]{cordes1986refractive}
Cordes, J., Pidwerbetsky, A., \& Lovelace, R. 1986, \apj, 310, 737

\bibitem[{Doeleman {et~al.}(2009)Doeleman, Agol, Backer, Baganoff, Bower,
  Broderick, Fabian, Fish, Gammie, Ho, {et~al.}}]{doeleman2009imaging}
Doeleman, S., Agol, E., Backer, D., {et~al.} 2009, Science White Paper
  submitted to the ASTRO2010 Decadal Review Panels (arXiv:0906.3899)

\bibitem[{{Falcke} {et~al.}(2000){Falcke}, {Melia}, \& {Agol}}]{Falcke_2000}
{Falcke}, H., {Melia}, F., \& {Agol}, E. 2000, \apjl, 528, L13

\bibitem[{Fish {et~al.}(2014)Fish, Johnson, Lu, Doeleman, Bouman, Zoran,
  Freeman, Psaltis, Narayan, Pankratius, {et~al.}}]{fish2014imaging}
Fish, V.~L., Johnson, M.~D., Lu, R.-S., {et~al.} 2014, \apj, 795, 134

\bibitem[{Fish {et~al.}(2016)Fish, Akiyama, Bouman, Chael, Johnson, Doeleman,
  Blackburn, Wardle, Freeman, \& the Event Horizon
  Telescope~Collaboration}]{galaxies4040054}
Fish, V.~L., Akiyama, K., Bouman, K.~L., {et~al.} 2016, Galaxies, 4

\bibitem[{Ghez {et~al.}(2008)Ghez, Salim, Weinberg, Lu, Do, Dunn, Matthews,
  Morris, Yelda, Becklin, {et~al.}}]{ghez2008measuring}
Ghez, A., Salim, S., Weinberg, N., {et~al.} 2008, \apj, 689, 1044

\bibitem[{Gillessen {et~al.}(2017)Gillessen, Plewa, Eisenhauer, Sari, Waisberg,
  Habibi, Pfuhl, George, Dexter, von Fellenberg,
  {et~al.}}]{gillessen2017update}
Gillessen, S., Plewa, P., Eisenhauer, F., {et~al.} 2017, \apj, 837, 30

\bibitem[{Goldreich \& Sridhar(1995)}]{goldreich1995toward}
Goldreich, P., \& Sridhar, S. 1995, \apj, 438, 763

\bibitem[{Goldreich \& Sridhar(2006)}]{goldreich2006folded}
---. 2006, \apjl, 640, L159

\bibitem[{Goodman \& Narayan(1989)}]{goodman1989shape}
Goodman, J., \& Narayan, R. 1989, \mnras, 238, 995

\bibitem[{{Gravity Collaboration} {et~al.}(2018){Gravity Collaboration},
  {Abuter}, {Amorim}, {Anugu}, {Baub{\"o}ck}, {Benisty}, {Berger}, {Blind},
  {Bonnet}, {Brandner}, {Buron}, {Collin}, {Chapron}, {Cl{\'e}net}, {Coud{\'e}
  Du Foresto}, {de Zeeuw}, {Deen}, {Delplancke-Str{\"o}bele}, {Dembet},
  {Dexter}, {Duvert}, {Eckart}, {Eisenhauer}, {Finger}, {F{\"o}rster
  Schreiber}, {F{\'e}dou}, {Garcia}, {Garcia Lopez}, {Gao}, {Gendron},
  {Genzel}, {Gillessen}, {Gordo}, {Habibi}, {Haubois}, {Haug}, {Hau{\ss}mann},
  {Henning}, {Hippler}, {Horrobin}, {Hubert}, {Hubin}, {Jimenez Rosales},
  {Jochum}, {Jocou}, {Kaufer}, {Kellner}, {Kendrew}, {Kervella}, {Kok},
  {Kulas}, {Lacour}, {Lapeyr{\`e}re}, {Lazareff}, {Le Bouquin}, {L{\'e}na},
  {Lippa}, {Lenzen}, {M{\'e}rand}, {M{\"u}ler}, {Neumann}, {Ott}, {Palanca},
  {Paumard}, {Pasquini}, {Perraut}, {Perrin}, {Pfuhl}, {Plewa}, {Rabien},
  {Ram{\'{\i}}rez}, {Ramos}, {Rau}, {Rodr{\'{\i}}guez-Coira}, {Rohloff},
  {Rousset}, {Sanchez-Bermudez}, {Scheithauer}, {Sch{\"o}ller}, {Schuler},
  {Spyromilio}, {Straub}, {Straubmeier}, {Sturm}, {Tacconi}, {Tristram},
  {Vincent}, {von Fellenberg}, {Wank}, {Waisberg}, {Widmann}, {Wieprecht},
  {Wiest}, {Wiezorrek}, {Woillez}, {Yazici}, {Ziegler}, \&
  {Zins}}]{2018A&A...615L..15G}
{Gravity Collaboration}, {Abuter}, R., {Amorim}, A., {et~al.} 2018, \aap, 615,
  L15

\bibitem[{Gwinn {et~al.}(2014)Gwinn, Kovalev, Johnson, \&
  Soglasnov}]{gwinn2014discovery}
Gwinn, C., Kovalev, Y.~Y., Johnson, M., \& Soglasnov, V. 2014, \apjl, 794, L14

\bibitem[{Hawking(1972)}]{hawking1972black}
Hawking, S.~W. 1972, CMaPh, 25, 152

\bibitem[{Israel(1967)}]{israel1967event}
Israel, W. 1967, PhRv, 164, 1776

\bibitem[{Israel(1968)}]{israel1968event}
---. 1968, CMaPh, 8, 245

\bibitem[{Johannsen \& Psaltis(2010)}]{johannsen2010testing}
Johannsen, T., \& Psaltis, D. 2010, \apj, 718, 446

\bibitem[{Johnson(2016)}]{johnson2016stochastic}
Johnson, M.~D. 2016, \apj, 833, 74

\bibitem[{Johnson \& Gwinn(2015)}]{johnson2015theory}
Johnson, M.~D., \& Gwinn, C.~R. 2015, \apj, 805, 180

\bibitem[{Johnson \& Narayan(2016)}]{johnson2016optics}
Johnson, M.~D., \& Narayan, R. 2016, \apj, 826, 170

\bibitem[{{Johnson} {et~al.}(2017){Johnson}, {Bouman}, {Blackburn}, {Chael},
  {Rosen}, {Shiokawa}, {Roelofs}, {Akiyama}, {Fish}, \&
  {Doeleman}}]{Johnson_2017}
{Johnson}, M.~D., {Bouman}, K.~L., {Blackburn}, L., {et~al.} 2017, \apj, 850,
  172

\bibitem[{Johnson {et~al.}(2018)Johnson, Narayan, Psaltis, Blackburn, Kovalev,
  Gwinn, Zhao, Bower, Moran, Kino, {et~al.}}]{johnson2018scattering}
Johnson, M.~D., Narayan, R., Psaltis, D., {et~al.} 2018, \apj, 865, 104

\bibitem[{{Kuramochi} {et~al.}(2018){Kuramochi}, {Akiyama}, {Ikeda}, {Tazaki},
  {Fish}, {Pu}, {Asada}, \& {Honma}}]{Kuramochi_2018}
{Kuramochi}, K., {Akiyama}, K., {Ikeda}, S., {et~al.} 2018, \apj, 858, 56

\bibitem[{{Lu} {et~al.}(2016){Lu}, {Roelofs}, {Fish}, {Shiokawa}, {Doeleman},
  {Gammie}, {Falcke}, {Krichbaum}, \& {Zensus}}]{Lu_2016}
{Lu}, R., {Roelofs}, F., {Fish}, V.~L., {et~al.} 2016, \apj, 817, 173

\bibitem[{Luminet(1979)}]{luminet1979image}
Luminet, J.-P. 1979, A\&A, 75, 228

\bibitem[{Narayan(1992)}]{narayan1992physics}
Narayan, R. 1992, PSPTA, 341, 151

\bibitem[{Narayan \& Goodman(1989)}]{narayan1989shape}
Narayan, R., \& Goodman, J. 1989, \mnras, 238, 963

\bibitem[{Nixon \& Aguado(2012)}]{nixon2012feature}
Nixon, M.~S., \& Aguado, A.~S. 2012, Feature extraction \& image processing for
  computer vision (Academic Press)

\bibitem[{Ortiz-Le{\'o}n {et~al.}(2016)Ortiz-Le{\'o}n, Johnson, Doeleman,
  Blackburn, Fish, Loinard, Reid, Castillo, Chael, Hern{\'a}ndez-G{\'o}mez,
  {et~al.}}]{ortiz2016intrinsic}
Ortiz-Le{\'o}n, G.~N., Johnson, M.~D., Doeleman, S.~S., {et~al.} 2016, \apj,
  824, 40

\bibitem[{Psaltis {et~al.}(2018)Psaltis, Johnson, Narayan, Medeiros, Blackburn,
  \& Bower}]{psaltis2018model}
Psaltis, D., Johnson, M., Narayan, R., {et~al.} 2018, \apj, submitted
  (arXiv:1805.01242)

\bibitem[{Psaltis {et~al.}(2015)Psaltis, {\"O}zel, Chan, \&
  Marrone}]{psaltis2015general}
Psaltis, D., {\"O}zel, F., Chan, C.-K., \& Marrone, D.~P. 2015, \apj, 814, 115

\bibitem[{Reid {et~al.}(2014)Reid, Menten, Brunthaler, Zheng, Dame, Xu, Wu,
  Zhang, Sanna, Sato, {et~al.}}]{reid2014trigonometric}
Reid, M., Menten, K., Brunthaler, A., {et~al.} 2014, \apj, 783, 130

\bibitem[{Rickett(1977)}]{rickett1977interstellar}
Rickett, B.~J. 1977, \araa, 15, 479

\bibitem[{Rickett(1990)}]{rickett1990radio}
---. 1990, ARA\&A, 28, 561

\bibitem[{Robinson(1975)}]{robinson1975uniqueness}
Robinson, D.~C. 1975, \prl, 34, 905

\bibitem[{Romani {et~al.}(1986)Romani, Narayan, \&
  Blandford}]{romani1986refractive}
Romani, R.~W., Narayan, R., \& Blandford, R. 1986, \mnras, 220, 19

\bibitem[{{S{\c a}dowski} {et~al.}(2014){S{\c a}dowski}, {Narayan}, {McKinney},
  \& {Tchekhovskoy}}]{Sadowski_2014}
{S{\c a}dowski}, A., {Narayan}, R., {McKinney}, J.~C., \& {Tchekhovskoy}, A.
  2014, \mnras, 439, 503

\bibitem[{{S{\c a}dowski} {et~al.}(2013){S{\c a}dowski}, {Narayan},
  {Tchekhovskoy}, \& {Zhu}}]{Sadowski_2013}
{S{\c a}dowski}, A., {Narayan}, R., {Tchekhovskoy}, A., \& {Zhu}, Y. 2013,
  \mnras, 429, 3533

\bibitem[{{S{\c a}dowski} {et~al.}(2017){S{\c a}dowski}, {Wielgus}, {Narayan},
  {Abarca}, {McKinney}, \& {Chael}}]{Sadowski_2017}
{S{\c a}dowski}, A., {Wielgus}, M., {Narayan}, R., {et~al.} 2017, \mnras, 466,
  705

\bibitem[{Schekochihin {et~al.}(2004)Schekochihin, Cowley, Taylor, Maron, \&
  McWilliams}]{schekochihin2004simulations}
Schekochihin, A.~A., Cowley, S.~C., Taylor, S.~F., Maron, J.~L., \& McWilliams,
  J.~C. 2004, \apj, 612, 276

\bibitem[{Scheuer(1968)}]{scheuer1968amplitude}
Scheuer, P. 1968, Natur, 218, 920

\bibitem[{Sch{\"o}del {et~al.}(2003)Sch{\"o}del, Ott, Genzel, Eckart, Mouawad,
  \& Alexander}]{schodel2003stellar}
Sch{\"o}del, R., Ott, T., Genzel, R., {et~al.} 2003, \apj, 596, 1015

\bibitem[{{Spitler} {et~al.}(2014){Spitler}, {Lee}, {Eatough}, {Kramer},
  {Karuppusamy}, {Bassa}, {Cognard}, {Desvignes}, {Lyne}, {Stappers}, {Bower},
  {Cordes}, {Champion}, \& {Falcke}}]{Spitler_2014}
{Spitler}, L.~G., {Lee}, K.~J., {Eatough}, R.~P., {et~al.} 2014, \apjl, 780, L3

\bibitem[{{Tatarskii}(1971)}]{tatarskii1971effects}
{Tatarskii}, V.~I. 1971, {The effects of the turbulent atmosphere on wave
  propagation}

\bibitem[{{Thompson} {et~al.}(1986){Thompson}, {Moran}, \&
  {Swenson}}]{thompson2001interferometry}
{Thompson}, A.~R., {Moran}, J.~M., \& {Swenson}, G.~W. 1986, {Interferometry
  and synthesis in radio astronomy}

\end{thebibliography}

\end{document}